\documentclass[prd,aps,twocolumn,a4paper,showkeys,nofootinbib,floatfix]{revtex4-1}

\usepackage{graphicx,psfrag}
\usepackage{mathrsfs}
\usepackage{amsmath,amsfonts,amssymb}
\usepackage{multirow}
\usepackage{comment}
\usepackage{ulem} % \sout{} remove in final version
\usepackage{hyperref}
\usepackage{enumitem}
\usepackage{morefloats}
\usepackage{bm}		% package for bold fonting greek letters in math mode

\newcommand{\be}{\begin{equation}}
\newcommand{\ee}{\end{equation}}
\newcommand{\bel}{\begin{align}}
\newcommand{\eel}{\end{align}}

\def\Msun{{M_{\odot}}}

\def\GMc2{{\rm G M_{\odot} c^{-2}}}

\def\kt2{\kappa^\text{T}_2}

\def\kt2{\kappa^\text{T}_2}

\newcommand{\rLR}{r_{\rm LR}}

\newcommand{\TEOB}[1]{\texttt{TEOBResumS{#1}}}
\newcommand\gsftides[1]{{\rm GSF{#1}}$^{(+)}${\rm PN}$^{(-)}$}

\usepackage{color}
\definecolor{cyan}{rgb}{0,0.9,0.9}
\definecolor{orange}{rgb}{0.9,0.5,0}
\definecolor{magenta}{rgb}{1,0,1}
\definecolor{purple}{rgb}{0.8,0.4,0.8}
\definecolor{gray}{rgb}{0.5,0.5,0.5}

\begin{document}

\title{Analytically improved and numerical-relativity informed effective-one-body model\\ for coalescing binary neutron stars}

\author{Rossella \surname{Gamba}${}^{1}$}
\author{Matteo \surname{Breschi}${}^{1,2}$}
\author{Sebastiano \surname{Bernuzzi}${}^{1}$}
\author{Alessandro \surname{Nagar}${}^{3,4,5}$}
\author{William \surname{Cook}${}^{1}$}
\author{Georgios \surname{Doulis}${}^{1}$}
\author{Francesco \surname{Fabbri}${}^{1}$}
\author{Néstor \surname{Ortiz}${}^{6}$}
\author{Amit \surname{Poudel}${}^{7}$}
\author{Alireza \surname{Rashti}${}^{8,9}$}
\author{Wolfgang \surname{Tichy}${}^{7}$}
\author{Maximiliano \surname{Ujevic}${}^{10}$}

\affiliation{${}^1$Theoretisch-Physikalisches Institut, Friedrich-Schiller-Universit{\"a}t Jena, 07743, Jena, Germany}  
\affiliation{${}^2$Theoretical and Scientific Data Science group, Scuola Internazionale Superiore di Studi Avanzati (SISSA), via Bonomea 265, 34136 Trieste Italy}
\affiliation{${}^3$Dipartimento di Fisica, Universit\`a di Torino, Torino, 10125, Italy}
\affiliation{${}^4$INFN sezione di Torino, Torino, 10125, Italy}
\affiliation{${}^5$Institut des Hautes Etudes Scientifiques, 35 Route de Chartres, Bures-sur-Yvette, 91440, France}
\affiliation{${}^6$Instituto de Ciencias Nucleares, Universidad Nacional Aut\'onoma de México, Circuito Exterior C.U., A.P. 70-543, México D.F. 04510, México}
\affiliation{${}^7$Department of Physics, Florida Atlantic University, Boca Raton, FL 33431, USA}
\affiliation{${}^8$Institute for Gravitation \& the Cosmos, The Pennsylvania State University, University Park, PA 16802, USA}
\affiliation{${}^9$Department of Physics, The Pennsylvania State University, University Park, PA 16802, USA}
\affiliation{${}^{10}$Centro de Ci\^encias Naturais e Humanas, Universidade Federal do ABC, Santo André 09210-170, SP, Brazil}
\date{\today}

\begin{abstract}
Gravitational wave astronomy pipelines rely on template waveform models for searches and parameter estimation purposes.
For coalescing binary neutron stars (BNS), such models need to accurately reproduce numerical relativity (NR) up to merger, 
in order to provide robust estimate of the stars' equation of state - dependent parameters.
In this work we present an improved version of the Effective One Body (EOB) model 
\TEOB{} for gravitational waves from BNS systems. Building upon recent post-Newtonian calculations, we include subleading 
order tidal terms in the waveform multipoles and EOB metric potentials, as well as add up to 5.5PN terms in the gyro-gravitomagnetic 
functions entering the spin-orbit sector of the model. In order to further improve the EOB-NR agreement in the last few orbital cycles before merger, 
we introduce next-to-quasicircular corrections in the waveform -- informed by a large number of BNS NR simulations -- and introduce a new NR-informed parameter entering the
tidal sector of our conservative dynamics.
The performance of our model is then validated against 14 new eccentricity reduced simulations of unequal mass, spinning
binaries with varying equation of state. A time-domain phasing analysis and mismatch computations demonstrate
that the new model overall improves over the previous version of \TEOB{}.
Finally, we present a closed-form frequency domain representation of the (tidal) amplitude and phase of the new \TEOB{} model. 
This representation accounts for mass-ratio, aligned spin and (resummed) spin-quadrupole effects in the tidal phase and 
-- within the calibration region -- it is faithful to the original model.
\end{abstract}

\maketitle

\section{Introduction}

Gravitational Waves (GWs) from coalescing binary neutron stars (BNS)
carry information on the stars' internal structure and composition, i.e. their equation of state (EOS)~\cite{Flanagan:2007ix,Hinderer:2007mb}. Such information is mainly, but not exclusively,
encoded in the tidal parameters of the stars themselves, which describe
the tidal response of a body due to the external gravitational field of its companion~\cite{Binnington:2009bb,Damour:2009vw}.
Precise measurements of the tidal parameters of neutron stars 
are a key science goal for current and next generation detectors~\cite{Radice:2017lry, Raithel:2019uzi, Capano:2019eae, Breschi:2021xrx, Puecher:2022oiz}.
As such, it is fundamental that waveform models -- which are extensively
employed during parameter estimation (PE) to extract the source properties from the data -- give faithful representations of the entire coalescence up to 
merger. Comparisons between current state of the art models (also called approximants, as they provide approximate solutions to the general relativistic two body problem) and Numerical relativity (NR) simulations, 
however, highlight that approximants do not correctly reproduce NR waveforms in the last stages of the inspiral, when matter contributions are best measured and their impact is the largest \cite{Gamba:2020wgg, Dudi:2021wcf, Gamba:2022mgx}. 
Numerous studies have shown that the imperfect modeling of matter effects will have large repercussions on PE with next-generation (XG) detectors, 
with waveform systematics that could potentially already be relevant for signals detected in the next LIGO-Virgo-Kagra observing run, O4.

% discussion on current state of the art models for BNS:
% phenom_Tidal
% SEOB
% TEOB

In this paper we improve the tidal sector of the effective one body (EOB) model \TEOB~\cite{Akcay:2018yyh,Nagar:2018plt,Nagar:2020pcj,Gamba:2021ydi}, and provide a phenomenological representation of it for spinning BNS systems.
The improvements concern (i) the computation and inclusion of additional higher order analytical information in the metric and waveform multipoles, and 
(ii) the inclusion of NR information via next to quasi-circular (NQC) parameters and through an additional NR-calibrated parameter, 
which enters the tidal part of the radial metric potential. For equal mass binaries, this parameter is clearly correlated with the effective tidal parameter of the system, 
and given a large enough number of high-quality simulations can be fit directly to NR. 

The paper is structured as follows. In Sec.~\ref{sec:eobmodel} we describe the structure of our model and the analytical improvements that we consider. 
In Sec.~\ref{sec:nrinfo} we discuss the NR information added and show comparisons between our model and few high resolution simulations, used 
as calibration set. 
In Sec.~\ref{sec:nrcomp} we validate the resulting model via additional comparisons against a set of new eccentricity reduced NR simulations characterized by rather extreme parameters, such as unequal masses and spins. In Sec.~\ref{sec:tidal} we construct a phenomenological representation of our model in the frequency domain, 
which can easily be added to any point mass binary black hole (BBH) model to reproduce the tidal sector of \TEOB{}.
Finally, in Sec.~\ref{sec:conc} we summarize and discuss our results.

Unless otherwise specified, we work in geometric units with $G=c=1$.
We label with $A, B$ the two stars and denote with $M_A, M_B$ the masses of the component stars, $M = M_A+M_B$ is the total mass of the system, 
$q\geq1$ is its mass ratio, $X_A = M_A/M$ and $X_B = M_B/M$ are the mass fractions, $\chi_A$ and $\chi_B$ are the (dimensionless) $z$ components of the 
spin angular momenta $S_A, S_B$, $\nu = X_A X_B$ is the symmetric mass ratio.

We further denote the electric-type, $\ell$-th multipole coefficient of the body $A$ as:
\begin{equation}
\mu_\ell^{A} = 2 \frac{k_{\ell}^A}{(2 \ell - 1)!!} R^{2 \ell + 1}_{A} \, ,
\end{equation}
where $k_{\ell}^A$ is the $\ell$-th electric Love number of body $A$, and $R_A$ is the radius of body $A$. 
The dimensionless tidal parameters $\Lambda_{\ell}$ are defined as
\begin{equation}
\Lambda_{\ell}^A = \frac{\mu^A_\ell}{M_A^{2 \ell +1}}  =  \frac{2}{(2 \ell +1)!!} k_{\ell} \mathcal{C}_A^{-(2 \ell +1)} \, ,
\end{equation}
where $\mathcal{C}_A=R_A/M_A$ is the compactness of body $A$.
Note that the usual NS parameter $\Lambda_A$ employed in GW data analysis refers to the $\ell = 2$ dimensionless tidal parameter above.
The effective tidal parameter $\tilde\Lambda$ is then obtained from the component stars' tidal parameters via
\be
\tilde\Lambda = \frac{16}{13} (X_A + 12 X_B)X_A^4 \Lambda_A + (A \leftrightarrow B) \, .
\ee

The EOB electric tidal coefficients, instead, are usually denoted as $\kappa^{(\ell +)}$. They are related to the tidal Love number $k^A_{\ell}$ through
\begin{eqnarray}
\kappa_{A}^{(\ell +)} &= 2 k_{\ell} \frac{X_B}{X_A} \Bigl(\frac{R_A}{M}  \Bigr)^{2 \ell +1} = 2 k_{\ell} \frac{X_B}{X_A} \mathcal{C}_A^{2 \ell +1} \, .
%G \mu_\ell^{A} &= \kappa^{A}_{\ell, +} \frac{X_A}{X_B} \frac{G^{2 \ell +1}}{(2 \ell +1)!!} \Bigl( \frac{M_A}{c^2} \Bigr)^{2 \ell +1}
\end{eqnarray}
Similarly, for the magnetic-type coefficients we denote with $j^A_\ell$ the magnetic-tipe love number, and define
the $\ell$-th multipole coefficient $\sigma_\ell^{A} $, the dimensionless tidal parameter $\Sigma_{\ell}^A$ and the EOB tidal coefficient $\kappa_A^{(\ell -)}$ as:
\begin{eqnarray}
\sigma_\ell^{A}     =& \frac{ (\ell -1) j_{\ell}^A}{4(\ell +2)(2 \ell - 1)!!} R^{2 \ell + 1}_{A} \, ,\\
\Sigma_{\ell}^A     =& \frac{\sigma_\ell^{A}}{M_A^{2 \ell +1}} = \frac{ (\ell -1) j_{\ell}^A}{4(\ell +2)(2 \ell - 1)!!} \mathcal{C}_A^{-(2 \ell +1)} \, ,\\
\kappa^{(\ell -)}_A =& \frac{1}{2} \frac{X_B}{X_A} j^A_\ell \Bigl( \frac{R_A}{M} \Bigr)^{2 \ell +1} = \frac{1}{2} \frac{X_B}{X_A} j^A_\ell \mathcal{C}_A^{2 \ell +1} \, .
\end{eqnarray}

\section{EOB Tidal model}
\label{sec:eobmodel}

The \TEOB ~model is a state-of-the-art EOB waveform model for generic-spin coalescing compact binaries \cite{Damour:2014sva, Nagar:2015xqa, Nagar:2018zoe, Nagar:2019wds, Nagar:2020pcj, Riemenschneider:2021ppj, Akcay:2020qrj, Gamba:2021ydi}.
As all EOB models, it is made up of three main building blocks, in principle separate from one another: a Hamiltonian describing the conservative motion, a radiation reaction force which accounts for energy and momentum losses and a prescription for the waveform at infinity. 
Below, we discuss the model used for BNS systems, highlighting improvements
and differences with respect to the previous model of \cite{Akcay:2018yyh}, here dubbed \gsftides{3}, in each of the fundamental blocks mentioned.

\subsection{Hamiltonian}

The EOB Hamiltonian is given by~\cite{Buonanno:1998gg, Damour:2000we, Damour:2015isa}
\be
\label{eq:HEOB}
\hat{H}_{\rm EOB} = \frac{1}{\nu}\sqrt{1+2\nu(\hat{H}_{\rm eff}-1)} \, ,
\ee
with
\be
\label{eq:Heff_orb}
\hat{H}_{\rm eff} = \hat{H}_{\rm eff}^{\rm orb} = \sqrt{p_{r*}^2 + A(r)\Bigl(1+ \frac{p_{\varphi}^2}{r^2}\Bigr)} \,.
\ee
in the case of nonspinning systems.
Here, $p_\varphi$ is the orbital angular momentum, $p_{r*}$ is the tortoise coordinate associated to the radial momentum $p_r$ given by $p_{r*} = \sqrt{A(r)/B(r)}  p_r$, and $A(r)$, $B(r)$ are the EOB metric potentials.
Within \TEOB, this Hamiltonian includes point-mass post-Newtonian (PN) analytical information up to 4PN in $A(r)$, and up to 3PN in the auxiliary potential $D(r)$, defined such that $A(r) B(r) = D(r)$.
Most of the analytical information described above is resummed via Padé approximants, to achieve robustness in the strong field, and further augmented by one pseudo-5PN coefficient which enters $A(r)$, $a_6^c$, calibrated to NR simulations of coalescing BBHs~\cite{Nagar:2018zoe}. 

\subsubsection{Spin}
Spins are included in the model in the Damour-Jaranowsky-Schaefer (DJS) gauge \cite{Damour:2008qf, Nagar:2011fx} via a modification of the effective Hamiltonian. Spin-orbit terms are accounted by two gyro-gravitomagnetic coefficients, $\hat{G}_{\hat{S}}$ and $\hat{G}_{\hat{S}_*}$, which are inverse-resummed and added to Eq.~\eqref{eq:Heff_orb}:
\be 
\hat{H}_{\rm eff} = \hat{H}_{\rm eff}^{\rm orb} + p_\varphi(\hat{G}_{S}\hat{S} + \hat{G}_{\hat{S}_*}\hat{S}_*) \, .
\ee 
$\hat{G}_{\hat{S}}$ and $\hat{G}_{\hat{S}_*}$ are currently (partially) known up to $5.5$ PN, 
i.e. to next-to-next-to-next-to-next-to-leading order (N4LO) \cite{Khalil:2021fpm}.
Here, we include and inverse-resum them up to this PN oder, improving on the previous model which only accounted for PN terms up to next-to-next to leading order (N2LO). %%; see App.~\ref{app:GSS} for their explicit expressions. 
Differently from BBH models, we do not employ any NR-informed coefficient in the gyro-gravitomagnetic terms.
Spin-squared terms are instead implemented to next-to-next-to-leading order (NNLO) in the model by 
replacing the EOB radial separation $r$ with the centrifugal radius $r_c$~\cite{Damour:2014sva}, defined as
\be
r^2_c = r^2 + \tilde{a}_Q^2\Bigl(1 + \frac{2}{r}\Bigr) + \frac{\delta\hat{a}_{\rm NLO}^2}{r} + \frac{\delta\hat{a}_{\rm NNLO}^2}{r^2}  \, ,
\ee
where $\tilde{a}_Q$ is an effective spin parameter that accounts for EOS-dependent spin-monupole interactions \cite{Nagar:2018plt} and reduces to the effective Kerr spin $\tilde{a}_0$ for BBH systems,
and beyond leading order (LO) spin squared contribution are included in $\delta\hat{a}_{\rm NLO}^2$ and  $\delta\hat{a}_{\rm NNLO}^2$, which can be read e.g. in \cite{Nagar:2018zoe, Nagar:2018plt}.

\subsubsection{Tidal effects}
Following the notation of \cite{Akcay:2018yyh}, electric $(+)$ and magnetic $(-)$ tidal interactions are generally included within the $A(r)$ and $B(r)$ potential as 
\be
A = A_0 + A_T^+ + A_T^- \, ,
\ee
\be
B = B_0 + B_T^+ + B_T^- \, .
\ee
where $A_0, B_0$ are the point mass BBH baselines of the metric potentials.
The tidal part of $A(r)$ is then typically further factorized as 
\be
\label{eq:A_tidal}
A_T^\pm(u) = \sum_{\ell \geq 2} A^{(\ell\pm)\rm LO}_A(u)\hat{A}_A^{\ell\pm}(u) + (A\leftrightarrow B)
\ee
where $u = 1/r$ is the inverse of the EOB radial coordinate, and the LO coefficients are straightforwardly given by
\be
A_A^{(\ell +)\rm LO}(u) = -\kappa_A^{(\ell +)}u^{2\ell +2} \, ,
\ee
and
\be
A_A^{(\ell -)\rm LO}(u) = -\kappa_A^{(\ell -)}u^{2\ell +3} \, .
\ee 
Within \TEOB, we consider LO electric contributions up to $\ell=8$ \cite{Godzieba:2021vnz}, extending the previous model of \cite{Akcay:2018yyh} that only included them up to $\ell=4$, and magnetic contributions up to $\ell=3$.

The higher order corrections to Eq.~(\ref{eq:A_tidal}) are known up to 2PN (NNLO)\cite{Bini:2012gu}. %%, and can be read from App.~\ref{app:APN}.
Notably however, rather than relying on the simple PN-expanded expressions above for 
$\hat{A}_A^{(2 +)}$ and $\hat{A}_A^{(3 +)}$, the \TEOB ~model implements a GSF-informed resummation of the potential:
\be
\hat{A}_A^{(\ell \pm)}(u) = \hat{A}_A^{(\ell \pm)\rm 0GSF} + X_A\hat{A}_A^{(\ell \pm) \rm 1GSF} + X_A^2 \hat{A}_A^{(\ell \pm) \rm 2GSF} \, .
\ee
The expressions for the GSF coefficients are collected in App.~\ref{app:APN}. 
Notably, all terms explicitly depend on the light ring radius $\rLR$, which determines the position of the pole in $A(u)$, and a ``GSF exponent" $p$, which appears in the 2GSF terms. The former used to be set by computing the adiabatic light ring of the $A(u)$ potential with 2PN tidal terms; the latter was fixed to $p=4$ independently of the binary parameters. 
Here, we make full use of the flexibility provided by the GSF resummation, fix $p=9/2$ and determine 
$r_{\rm LR}$ via comparisons to NR data (see Sec.~\ref{sec:nrinfo} below).

The $B(u)$ potential includes tidal effects to leading order in the electric and magnetic tidal parameters \cite{Vines:2010ca,Akcay:2018yyh}:
\begin{align}
B^+_T(u) &=  3(3-5\nu)(\kappa_A^{(2+)} + \kappa_B^{(2+)}) u^6 \, , \\
B^-_T(u) &=  5(\kappa_A^{(2-)} + \kappa_B^{(2-)}) u^6 \, .
\end{align}
The magnetic term is -- to the best of our knowledge -- presented here for the first time. 
Its computation used the results of Ref.~\cite{Henry:2019xhg}.
Following standard techniques, we perform a Legendre transformation of the harmonic center of mass 1PN 
Lagrangian of Ref.~\cite{Henry:2019xhg} proceeding order-by-order and
recovering an ``ADM-like'' Hamiltonian, which coincides with the usual ADM Hamiltonian in the point mass sector.
Via a canonical transformation, closely following the procedure outlined in \cite{Damour:2000we}, we express the Hamiltonian $\hat{H}_{\rm ADM-like}$ and $\hat{H}_{\rm eff}$ in the same 
set of coordinates.
We parameterize the 1PN generating function $G_{\rm 1PN}(\mathbf{q}, \mathbf{p}')$, where
$(\mathbf{q}, \mathbf{p})$ are the original (ADM-like) coordinates and $(\mathbf{q}', \mathbf{p}')$
are the desired EOB coordinates.
We then compute:
\begin{eqnarray}
\mathbf{q}' = \mathbf{q} + \frac{\partial G}{\partial \mathbf{p}'} \, ,\\
\mathbf{p} = \mathbf{p}' + \frac{\partial G}{\partial \mathbf{q}} \, ,
\end{eqnarray}
and compare:
\begin{equation}
\hat{H}_{\rm eff}^2(q',p') = \Bigl[ 1 + \hat{H}_{\rm ADM-like}(q,p) + \alpha_1 \hat{H}_{\rm ADM-like}(q,p)^2 \Bigr]^2 \, .
\end{equation}
the $A(u)$ and $B(u)$ potentials within $\hat{H}_{\rm eff}$ are expanded to 1PN, and the LO magnetic and electric terms in $B(u)$ are parameterized via two unknown 
coefficients $\chi_2$ and $\beta_2$, respectively.
We then find
\begin{eqnarray}
\beta_2 =& 3(3 - 5 \nu) \, ,\\
\chi_2 =&5 \, .
\end{eqnarray}
thus confirming the result of Vines et. al.~\cite{Vines:2010ca} and of Bini et. al.~\cite{Bini:2012gu}, and computing $\chi_2$ for the first time.
Notably, $B(u)$ is typically computed as the ratio of the EOB $D(u)$ BBH potential
and the $A(u)$ potential 
discussed above. Since $A(u)$ already contains tidal corrections, in order to obtain the correct $B(u)$ PN LO one needs to correct the ratio $D(u)/A(u)$ with an additional term $B'_T(u)$, so that in practice
\begin{align*}
B_T(u) &= \frac{D(u)}{A(u)} + B'_T(u) \, \\
B'_T(u) &= (\kappa_A^{(2+)} + \kappa_B^{(2+)})(8 -15\nu)u^6 + B^{-}_T(u) \, .
\end{align*}

\subsection{Waveform and radiation reaction}

The BBH EOB waveform is given by
\begin{equation}
h_{\ell m}^0 = c_{\ell + \epsilon}(\nu) h' {}_{\ell m}^{(N, \epsilon)} S^{(\epsilon)}h_{\ell m}^{\rm tail} f_{\ell m} \, = h' {}_{\ell m}^{(N, \epsilon)}\hat{h}^0_{\ell m} .
\end{equation}
where $h' {}_{\ell m}^{(N, \epsilon)}$ is a Newtonian prefactor,  $S^{(\epsilon)}$ is the source term, $h_{\ell m}^{\rm tail}$ includes tail contribitions and $f_{\ell m}$ resums the residual terms \cite{Damour:2008gu, Pan:2010hz, Damour:2014sva,Nagar:2016ayt,Messina:2018ghh,Nagar:2019wrt}.
Tidal contributions are included in the EOB multipoles $h_{\ell m}$ via a simple additive correction to the BBH baseline, 
$h_{\ell m} = h_{\ell m}^0 + h_{\ell m}^{\rm tidal}$.
The tidal part of the waveform multipoles does not follow the standard EOB factorization, and is instead simply 
given by:
\begin{eqnarray}
h_{\ell m}^{\rm tidal} =& h' {}_{\ell m}^{(N, \epsilon)}\hat{h}_{\ell m}^{\rm tidal}\, , \\
\hat{h}_{\ell m}^{\rm tidal} =& c_{\ell m}^{\rm LO}\kappa_A (1 + \beta_{\ell m}^1 x + \beta_{\ell m}^2 x^2) + (A \leftrightarrow B) \, .
\end{eqnarray}
where $c_{\ell m}^{\rm LO}$ and $\beta_{\ell m}^{n}$ parameterize the leading order and $n$-th PN corrections to the waveform amplitude.
These terms were previously known up to NLO \cite{Damour:2009wj, Vines:2010ca, Damour:2012yf}; 
here we exploit the results of Ref.~\cite{Henry:2020ski} to compute and include also the leading order correction to $h_{44}$ and $h_{42}$ and the subleading terms $\beta_{22}^2$ and $\beta_{21}^1$.

To obtain the desired terms, we first evaluate the multipolar fluxes $F_{\ell m}$, parameterized by the unknown coefficients, via
\begin{equation}
F_{\ell m} = F_{\ell m}^{(N, \epsilon)} |\hat{h}_{\ell m}|^2 = F_{\ell m}^{(N, \epsilon)} |\hat{h}_{\ell m}^0 + \hat{h}_{\ell m}^{\rm tidal}|^2 \, .
\end{equation}
The Newtonian flux prefactors $F_{\ell m}^{(N, \epsilon)}$ can be found in e.g. App.~A of~\cite{Messina:2018ghh}. 
Note however that, with respect to the definition given in Ref.~\cite{Messina:2018ghh}, 
the term $c_{\ell+\epsilon}(\nu)$ is here factored out and included directly in the definition of $\hat{h}_{\ell m}^0$.
We then compare our results to the multipolar fluxes computed in Ref.~\cite{Henry:2020ski}, and thus extract the desired parameters.
We find:

\begin{widetext}
\begin{eqnarray}
\beta_{22}^2 =& \frac{111970 X_A^7-414911 X_A^6+805952 X_A^5-728217 X_A^4+363195
   X_A^3-347284 X_A^2+153638 X_A-16366}{10584 X_A (2 X_A-3)} \, ,\\
\beta_{21}^1 =& \frac{-220 X_A^3-130 X_A^2+203 X_A+15}{126-168 X_A} \, ,\\
c_{44}^{LO} =&  2 (5 - 9 X_A + 6 X_A^2) \, ,\\
c_{42}^{LO} =&  3584 (5 - 9 X_A + 6 X_A^2) \, .
\end{eqnarray}
\end{widetext}

When specifying $\beta_{22}^2$ to the equal mass case ($X_A = X_B = 1/2$) we find $\beta_{22}^2 = 167/256 \sim 0.653$. 
This value is extremely close, but not exactly equal, to the $\beta_{22}^2$ estimated in Ref.~\cite{Henry:2020ski} when comparing their results 
with Ref.~\cite{Damour:2012yf}. We attribute this discrepancy to possible computation errors in Ref.~\cite{Damour:2012yf}, 
which was also found to be incorrect in the $7.5$PN tail term.
Crucially, with respect to previous versions of the model, we do not propagate the tail contribution $h_{\ell m}^{\rm tail}$ to the tidal amplitudes, 
in order to correctly recover the LO 6.5PN tail terms.
Given the new tidal terms contributing to the waveform amplitudes $\hat{h}_{\ell m}$, the radiation reaction force $\hat{\mathcal{F}}_\varphi$ is extended straightforwardly.

\subsection{Effect of the analytical information}

\begin{figure*}[t]
	\includegraphics[width=0.49\textwidth]{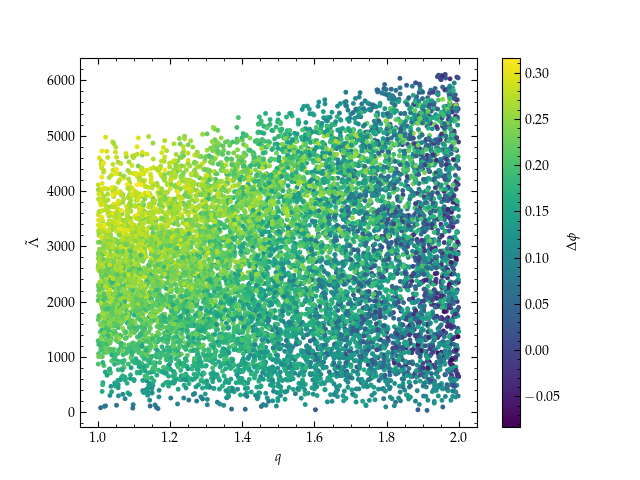}
	\includegraphics[width=0.49\textwidth]{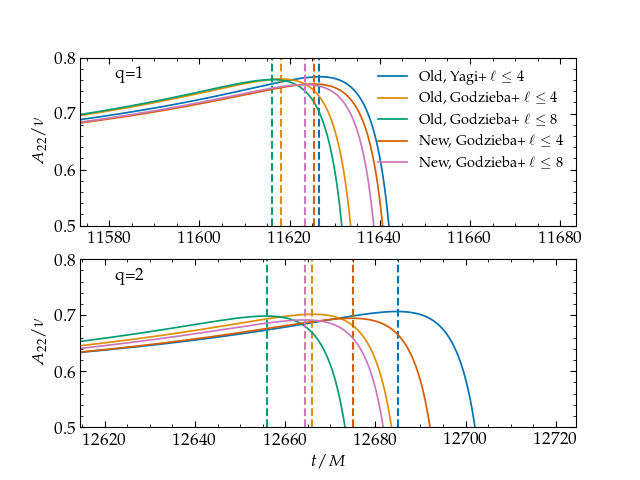}
	\caption{\label{fig:analytical} Impact of the new analytical information included in the model. 
	Left: phase differences $\Delta \phi = \phi^{\rm old} - \phi^{\rm new}$ between the model without (``old'') and with (``new'')
	the newly added information. The phase difference is positive over a large portion of the parameter space, indicating that 
	the newly computed terms have an overall repulsive effect on the dynamics, implying weaker tidal effects.
	Right: time evolution of the $(2,2)$ amplitude for two non-spinning BNSs with $\tilde{\Lambda}=2500$ and $q=1$ (top) and $q=2$ (bottom).
	The orange and pink curves displayed correspond to the ``old'' and ``new'' models compared in the left figure. The biggest difference
	between all the curves shown is caused by the number of multipolar tidal parameters $\Lambda_{\ell}$ considered, and the fits employed 
	to compute them. As the mass ratio increases, the tidal corrections added to the higher modes in the ``new'' models become more important,
	increasing the strength of tidal effects in this scenario.}
\end{figure*}

The left panel of Fig.~\ref{fig:analytical} shows the impact of the additional analytical information included in the Hamiltonian and the waveform
for $10^4$ systems with varying mass ratios, tidal parameters and spins. Such a difference is here quantified in terms of the phase difference at merger $\Delta \phi = \phi^{\rm old} - \phi^{\rm new}$, 
where ``new'' and ``old'' indicate the models with and without the extra terms discussed in the previous subsections. 
Note that, in both cases, $\ell > 2$ multipolar tidal parameters are estimated via the fits from Ref.~\cite{Godzieba:2020bbz, Godzieba:2021vnz}.
From this comparison it appears that: 
(i) the (absolute) values of the phase differences are smallest for binaries with small tidal parameters or larger mass ratios, 
and maximal for equal mass binaries with large tidal parameters; 
(ii) the phase difference is positive over a considerable portion of the parameter space, implying that the new tidal model 
prescribes weaker matter effects than the previous.
The facts observed above are, at first impact, rather puzzling: while it is expected for the models to differ the most for large values of 
$\tilde\Lambda$, it is not immediately clear why they should agree for high mass ratios, or -- conversely -- why they should differ 
the most for equal mass BNSs.

To further understand this picture, then, we consider two representative cases: an equal mass, non-spinning binary with $\tilde{\Lambda}=2500$, 
and a $q=2$ binary with the same effective tidal parameter and spins. 
For each, we also compute the waveform combining the ``new'' and ``old'' model with different fits for the multipolar tidal parameters 
$\Lambda_{\ell}$. The right panel of Fig.~\ref{fig:analytical} shows the $(2,2)$ amplitude evolution for the two systems considered.
Focusing on the equal mass case, the largest effect for this binary is given by choice of the fit for $\Lambda_{3,4}$. 
The inclusion of $\Lambda_{\ell}$ terms with $\ell \geq 5$, too, has a considerable effect on the waveform.
Contrasting models based on the same fits and with identical $\Lambda_{\ell}$ content, it appears that the tidal contributions to the $B(u)$ potential, 
the new tidal contributions to $h_{22}$ itself and the lack of propagation of $h_{\ell m}^{\rm tail}$ to the multipolar waveforms have an overall 
repulsive effect on the dynamics. When the mass ratio is increased the importance of higher modes also grows, 
and differences between models are attenuated. In this scenario, the ``new'' model with $\Lambda_{\ell \leq 8}$ is comparable to the ``old'' model
with  $\Lambda_{\ell \leq 4}$, which in the equal mass case was significantly more attractive. 

Qualitatively, moving from the model of Ref.~\cite{Akcay:2018yyh} (which used the fits of Ref.~\cite{Yagi:2013sva}), to a model which includes more analytical 
information and is based on the fits from Ref.~\cite{Godzieba:2020bbz, Godzieba:2021vnz} represents a step towards the right direction: 
previous EOB-NR comparisons highlighted how the models fail to capture the last few cycles before merger, with tidal effects not being attractive enough.
Quantitatively, however, the new contributions are relatively small: the dephasing at merger varies between $-0.05$ and $0.3$ rad.
The inclusion of the analytical information discussed in the
previous section to the conservative and radiative sectors of the model therefore
does not, unfortunately, provide corrections to the final waveform that are large enough to
fill the gap with NR simulations, as observed in e.g. \cite{Akcay:2018yyh,Gamba:2022mgx}.

\section{Numerically informing the model}
\label{sec:nrinfo}
\begin{figure*}[t]
	\includegraphics[width=0.49\textwidth]{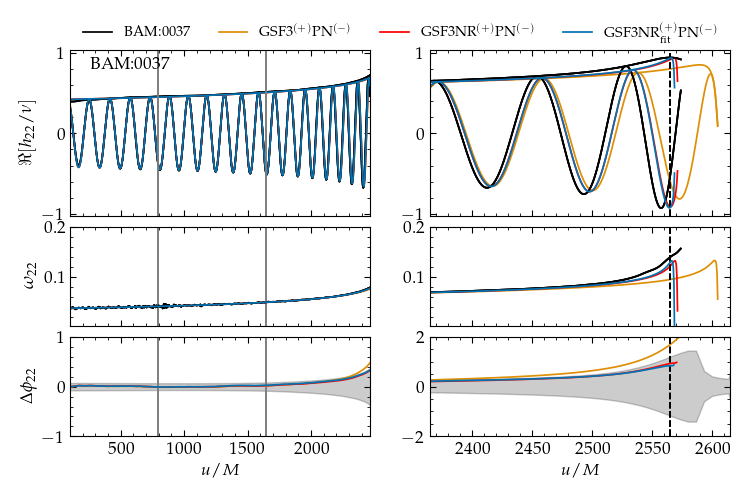}
	\includegraphics[width=0.49\textwidth]{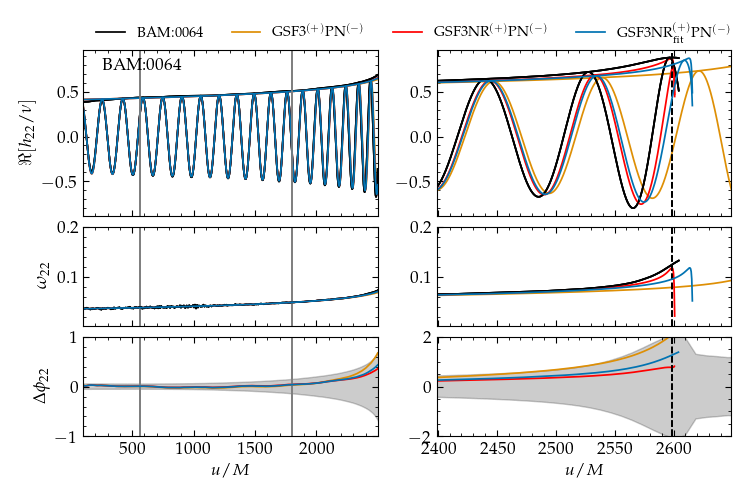}
	\includegraphics[width=0.49\textwidth]{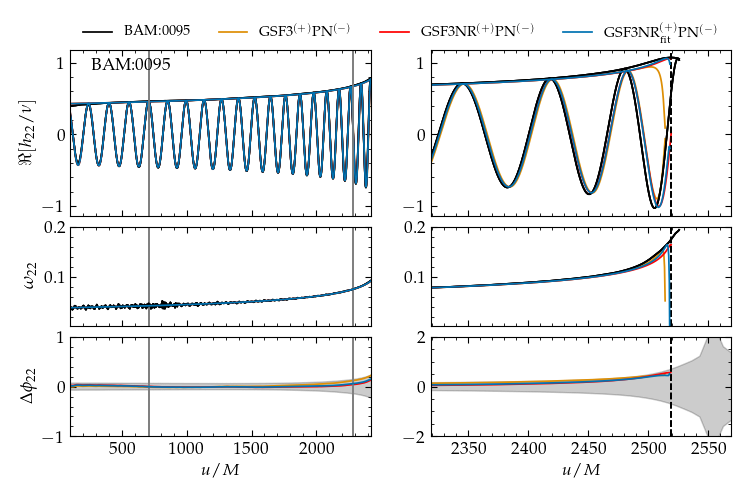}
	\includegraphics[width=0.49\textwidth]{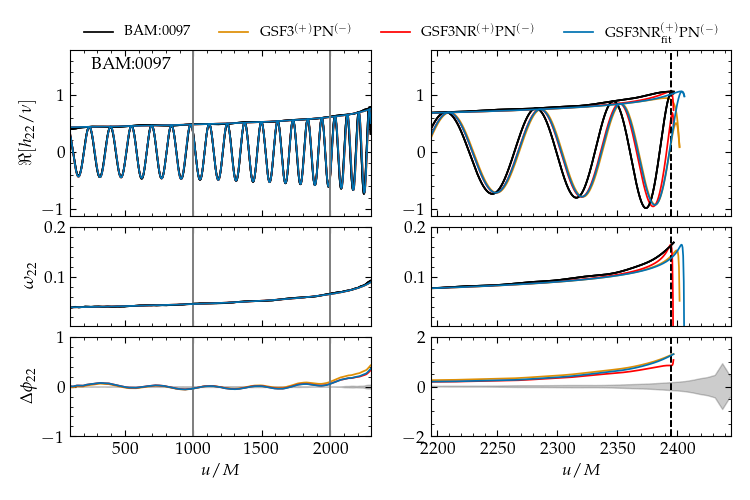}
	\caption{
		Comparison between different versions of the \TEOB~model and the four equal mass NR simulations
		employed to tune the tidal parameter $\alpha_T$. We display the real part of $h_{22}$ (top), the frequency evolution (middle)
		and the EOB/NR phase difference (bottom) as functions of the retarded time. The gray bands in the bottom panels 
		indicate the NR error, estimated as the sum in quadrature of resolution and extraction error. The vertical gray lines indicate
		the alignment region, and the vertical dashed line the time of merger. Orange lines correspond to the model of Ref.~\cite{Akcay:2018yyh}, \gsftides{3};
		red and blue lines to the new numerically-informed \TEOB model \gsftides{3NR} where $\alpha_T$ is estimated for each simulation (red) or via the fit of 
		Eq.~\ref{eq:alpha_fit}. 
		We find that \TEOB{}~reproduces NR up to merger in three out of four cases,
		but -- notably -- does not fully capture the NR frequency evolution over the final two orbits before merger. The addition of NR-informed
		parameters, however, clearly improves the EOB/NR phase agreement for the {\tt BAM:0037} and {\tt BAM:0064}
		simulations with respect to the model of Ref.~\cite{Akcay:2018yyh}.
	}
	\label{fig:bam_q1}
\end{figure*}

The need of employing some NR information to improve our model in the last few cycles before
merger appears inevitable. 
We proceed on two different fronts.
First, we include NQC corrections in our waveform, in order to ensure that
the expected NR-prescribed values of amplitude and frequency at merger are reached by our waveforms.
Then, 
we employ the flexibility of the tidal model described in Sec.~\ref{sec:eobmodel} provided by the light ring radius $r_{\rm LR}$ 
or the GSF-inspired exponent $p_{\rm GSF}$. These parameters effectively determine the strength of tidal interactions by shifting 
the position of the pole in the tidal potential ($r_{\rm LR}$) or the degree of the singularity ($p_{\rm GSF}$).

\subsection{NQC model for BNS systems}
NQCs, or ``next-to-quasicircular" corrections have been first introduced 
within the EOB formalism by Damour and Nagar in Ref.~\cite{Damour:2007xr}, 
and have proven to be fundamental in the creation of faithful inspiral-merger-ringdown BBH waveform models 
\cite{Nagar:2017jdw,Nagar:2018zoe,Riemenschneider:2021ppj,Albertini:2021tbt}. 
Here, we apply NQC corrections to BNS systems for the dominant $\ell=m=2$ mode, making use of the merger fits provided by
\cite{Breschi:2022xnc}.

We remind the reader that NQC corrections enter the factorized EOB waveform as a multiplicative term $\hat{h}_{\ell m}^{\rm NQC}$, which in the general case explicitly reads:
\be
\hat{h}_{\ell m}^{\rm NQC} = (1 + a_1 n_1 + a_2 n_2) e^{i(b_1 n_3 + b_2 n_4)} \, .
\ee
The numerical coefficients $a_1, a_2, b_1, b_2$ are to be determined from the values of NR amplitude and frequency at merger, while $n_1, n_2, n_3, n_4$ are the NQC basis functions. Consistently with the BBH case, we employ:
\begin{eqnarray}
n_1 &=  \frac{p_{r_*}^2}{r^2\Omega^2} ,\\
n_2 &=  \frac{\ddot{r}}{r \Omega^2} ,\\
n_3 &=  \frac{p_{r_*}}{r \Omega} ,\\
n_4 &=  n_3 r^2 \Omega^2 .
\end{eqnarray}
Following standard EOB techniques, the NQC coefficients $a_i, b_i$ are determined by imposing that the EOB amplitude $\hat{A}_{22}$, its time derivative $\dot{\hat{A}}_{22}$, the frequency $\hat{\omega}_{22}$ and the frequency time derivative $\dot{\hat{\omega}}_{22}$
extracted $3 M$ before the peak of the EOB orbital frequency are equal to the same NR quantities extracted at the NR NQC time $t_{\rm NQC}^{\rm NR}$, which is here identified with the merger, $t_{\rm NQC}^{\rm NR} = t_{\rm mrg}^{\rm NR}$.
\begin{eqnarray}
\hat{A}_{22}^{\rm EOB}(t_{\Omega^{\rm peak}}-3) &=  \hat{A}_{22}^{\rm NR}(t_{\rm mrg}) \, ,\\
\dot{\hat{A}}_{22}^{\rm EOB}(t_{\Omega^{\rm peak}}-3) &=  \dot{\hat{A}}_{22}^{\rm NR}(t_{\rm mrg}) = 0 \, ,\\
\hat{\omega}_{22}^{\rm EOB}(t_{\Omega^{\rm peak}}-3) &=  \hat{\omega}_{22}^{\rm NR}(t_{\rm mrg}) \, ,\\
\dot{\hat{\omega}}_{22}^{\rm EOB}(t_{\Omega^{\rm peak}}-3) &=  \dot{\hat{\omega}}_{22}^{\rm NR}(t_{\rm mrg}) \, ,\\
\end{eqnarray}
This choice of the extraction point is slightly different with respect to the BBH case, where $t_{\rm NQC}^{\rm NR} = t_{\rm mrg}^{\rm NR} + 2$, and is motivated by the desire of employing already existing fits to the desired NR quantities when they are specified at merger. In detail, we supplement the fits presented in Ref.~\cite{Breschi:2022xnc} for 
$\hat{A}_{22}^{\rm NR}(t_{\rm mrg})$ and $\hat{\omega}_{22}^{\rm NR}(t_{\rm mrg})$ with new ones for
$\dot{\hat{\omega}}_{22}^{\rm NR}(t_{\rm mrg})$ (see App.~\ref{app:domg_fit}).

\subsection{Light ring singularity}
The GSF-informed resummation introduced in \cite{Akcay:2018yyh} naturally introduced
a \textit{pole} in the denominator of the $\hat{A}^{\rm XGSF}$ terms, with $X=0,1,2$.
The precise location of this pole is however quite uncertain: 
the analytical GSF expressions placed the pole at $r=3$, where the BBH light ring is situated. However, clearly, given that we are not describing coalescing BHs, this value 
can be modified at will. One solution is to employ the light ring radius $r_{\rm LR}$ implied by the $A$ potential augmented by NNLO tidal effects. 
The value of $r_{\rm LR}$ so estimated is generally larger than the BBH light ring, effectively enhancing the tidal terms closer to the end of the evolution, in the last few orbital cycles.
Such an enhancement was shown to improve the agreement between NR and EOB, although
it was still not sufficient to reproduce NR within its error bars in most of the cases studied.

Here, we build on this approach by multiplying the light ring radius estimated from the NNLO $A$ potential by a parameter $\alpha$, 
to be deterimined case-by-case by comparing our model to high resolution NR simulations after alignment.
We start from the equal mass, non-spinning case, and consider the public $\tt CoRe$ simulations {\tt BAM:0037}, {\tt BAM:0064}, {\tt BAM:0095}, and {\tt BAM:0097}. 
All simulations show clear convergence properties, with the latter displaying evident $4^{\rm th}$ order convergence thanks to the Enthropy-Flux-Limited (EFL) 
method developed in \cite{GUERMOND2008801, Guercilena:2016fdl} and improved in \cite{Doulis:2022vkx}.
Given the multiple available resolutions and extraction radii associated to each simulation, 
we estimate the total error budget on the waveform phase as the sum in quadrature of the resolution and finite extraction error. 
The former is computed by evaluating the phase difference between the highest and second-highest resolutions; 
the latter by estimating the phase difference between the waveform extrapolated to infinity and extracted 
at the largest finite radius.

Comparisons between our $\alpha$-calibrated and uncalibrated EOB model with the NR
waveforms above are shown in Fig.~\ref{fig:bam_q1}. 
The calibrated model can reproduce all the simulations shown within their NR error, 
except for {\tt BAM:0097}. Notably, even when the phase error at merger is within the NR errorbars, 
the NQC-enhanced and NR-tuned EOB model still does not seem
to be able to fully capture the frequency evolution over the last two cycles. 
Nonetheless, the introduction of the $\alpha$ parameter noticeably improves the EOB/NR agreement 
for binaries with large effective tidal parameters, such as {\tt BAM:0037} and {\tt BAM:0064}. 
A trend in $\alpha$ is also easily identifiable: it is clear that to match the simulations with larger tidal parameters, 
larger values of $\alpha$ are necessary. 
Conversely, given that we set $p=9/2$, values of $\alpha < 1$ need to be chosen in order 
to obtain agreement with those simulations with $\kappa^T_2 \leq 100$.
Using the four simulations shown in Fig.~\ref{fig:bam_q1} we fit
\be
\label{eq:alpha_fit}
\alpha = \frac{a_1 \kappa_2^T}{1 + a_2 \kappa_2^T} \, ,
\ee
finding $a_1 = 0.06306$ and $a_2 = 0.05732$. 

Beyond our calibration set, we determine $\alpha$ for $\sim 10$ additional lower-resolution non-spinning 
simulations and compare the extracted values with the $\alpha$ computed via Eq.~\ref{eq:alpha_fit}. 
This investigation indicates that the trend we identified is maintained when considering other equal-mass 
configurations, with the largest $\Delta\alpha/\alpha$ amounting to about $13\%$.
Concerning unequal mass and spinning terms, instead, we do not attempt to fit these contributions to $\alpha$ 
due to the large uncertainties affecting the simulations. 
However, we mention that even in this scenario by modifying $\alpha$ it is usually possible to 
reproduce the NR waveforms within their estimated error bands.

\section{EOB/NR comparisons}
\label{sec:nrcomp}

\subsection{New simulations}

\begin{table}[t]
	% \resizebox{\textwidth}{!}{
	  \centering
	\begin{tabular}{ccccccccccc}
	\hline\hline
	ID & EOS & $M\omega_0$ & $M$ & $q$ & $\Lambda^A_2$ & $\Lambda^B_2$ & $\chi_A$ & $\chi_B$ & $\bar{\mathcal{F}}_{old}$ & $\bar{\mathcal{F}}_{new}$ \\
	\hline 
	ER01 & H4  & $0.0373$ & $2.77$ & $1.02$ & $888$ & $1007$ & $0.03$ & $0.07$ & $0.0119$ & $0.0039$\\
	ER02 & H4  & $0.0339$ & $2.6$ & $1.26$ & $719$ & $2789$ & $0.04$ & $0.13$ & $0.0225$ & $0.0045$\\
	ER03 & H4  & $0.0339$ & $2.6$ & $1.26$ & $718$ & $2794$ & $0.11$ & $0.05$ & $0.0219$ & $0.009$\\
	ER04 & H4  & $0.0373$ & $2.77$ & $1.02$ & $887$ & $1008$ & $0.06$ & $0.03$ & $0.0132$ & $0.0032$\\
	ER05 & H4  & $0.0382$ & $2.82$ & $1.06$ & $718$ & $1008$ & $0.11$ & $0.05$ & $0.008$ & $0.0024$\\
	ER06 & MS1b  & $0.0369$ & $2.77$ & $1.02$ & $1268$ & $1420$ & $0.1$ & $0.03$ & $0.0059$ & $0.0169$\\
	ER10 & MS1b  & $0.0347$ & $2.65$ & $1.21$ & $1048$ & $2854$ & $0.03$ & $0.14$ & $0.0339$ & $0.0072$\\
	ER11 & MS1b  & $0.0379$ & $2.82$ & $1.06$ & $1048$ & $1418$ & $0.03$ & $0.1$ & $0.0045$ & $0.0267$\\
	ER12 & MS1b  & $0.0379$ & $2.82$ & $1.06$ & $1048$ & $1417$ & $0.03$ & $0.13$ & $0.0203$ & $0.0014$\\
	ER13 & MS1b  & $0.0378$ & $2.82$ & $1.06$ & $1047$ & $1420$ & $0.09$ & $0.03$ & $0.018$ & $0.0044$\\
	ER14 & MS1b  & $0.0379$ & $2.82$ & $1.06$ & $1046$ & $1420$ & $0.12$ & $0.03$ & $0.0164$ & $0.0026$\\
	ER15 & H4  & $0.0382$ & $2.82$ & $1.05$ & $719$ & $1006$ & $0.04$ & $0.11$ & $0.0075$ & $0.0019$\\
	ER17 & H4  & $0.0373$ & $2.77$ & $1.02$ & $888$ & $1006$ & $0.05$ & $0.11$ & $0.0109$ & $0.0068$\\
	ER18 & H4  & $0.0373$ & $2.77$ & $1.02$ & $886$ & $1008$ & $0.11$ & $0.05$ & $0.0128$ & $0.0054$\\
	\hline
	\end{tabular}
	% }
	\caption{New spinning, unequal mass NR simulations used to test \TEOB{}. The new model has lower unfaithfulness that the previous one in all but two cases.}
	\label{tab:nr_sims}
\end{table}

We present 14 new simulations of unequal mass, spinning binaries. 
The initial data for these simulations has been computed using the SGRID library~\cite{Tichy:2011gw,Tichy:2012rp,Tichy:2019ouu}, and eccentricity reduction
has been performed in order to minimize residual spurious artifacts in the waveform~\cite{Moldenhauer:2014yaa,Dietrich:2015pxa}.
The constraint-satisfying data is then evolved with the {\tt BAM} code \cite{Bruegmann:1996kz,Brugmann:2008zz,Thierfelder:2011yi}.
All simulations are run at multiple resolutions, with 64, 96, 128 or 72, 108, 144 points per directions in the
finest (moving) mesh refinement level covering the individual NSs and using a high-order hydrodynamics scheme \cite{Bernuzzi:2016pie}. 
More details on the simulations are given in Tab.~\ref{tab:nr_sims} and in Appendix~\ref{app:NR}.

\subsection{Time-domain phasing}

\begin{figure*}
	\includegraphics[width=0.4\textwidth]{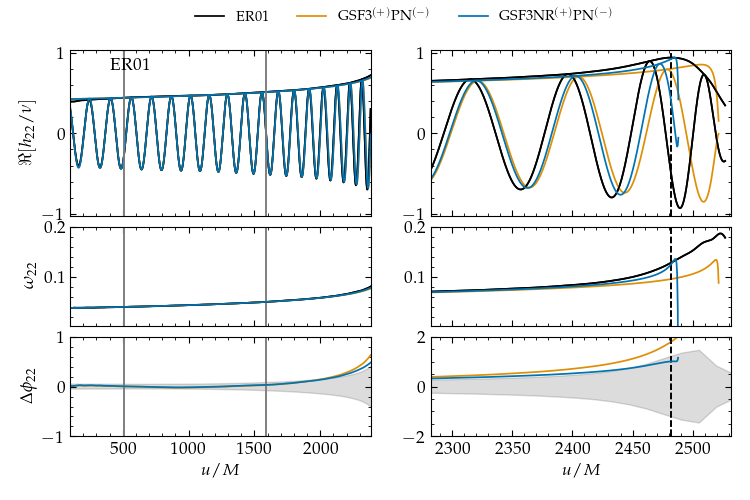}
	\includegraphics[width=0.4\textwidth]{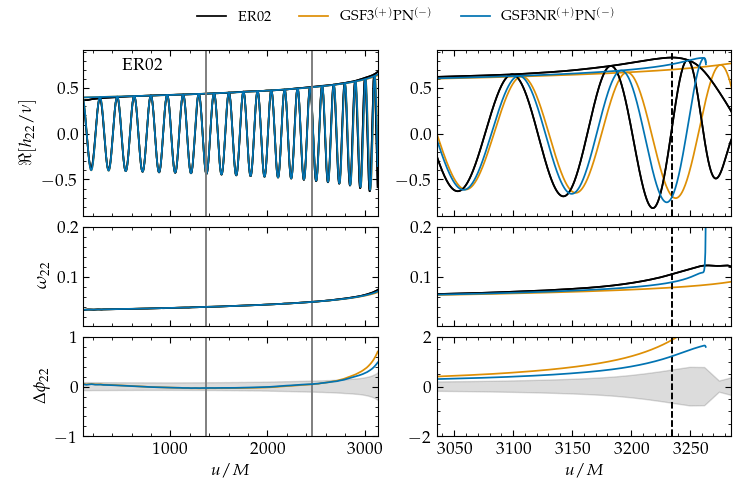}
	\includegraphics[width=0.4\textwidth]{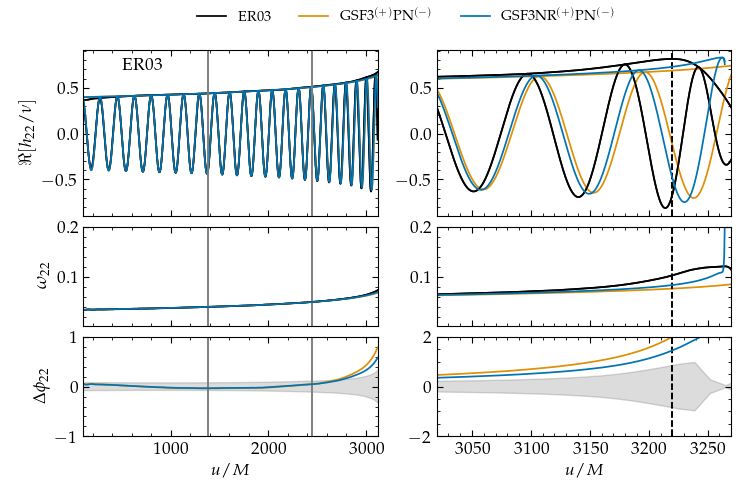}
	\includegraphics[width=0.4\textwidth]{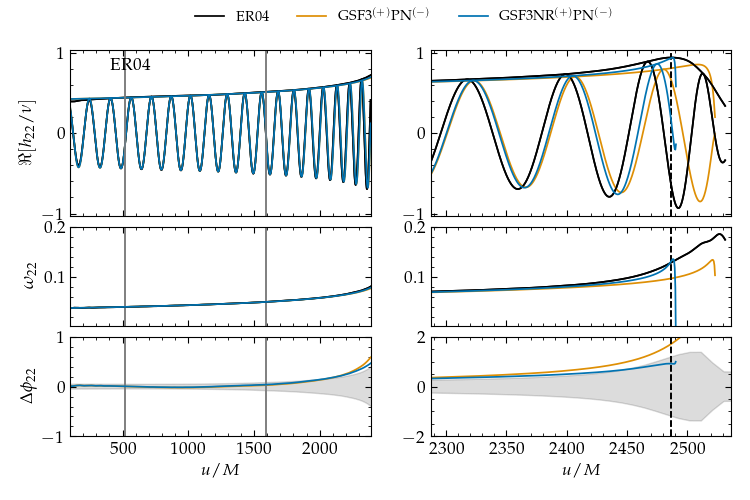}
	\includegraphics[width=0.4\textwidth]{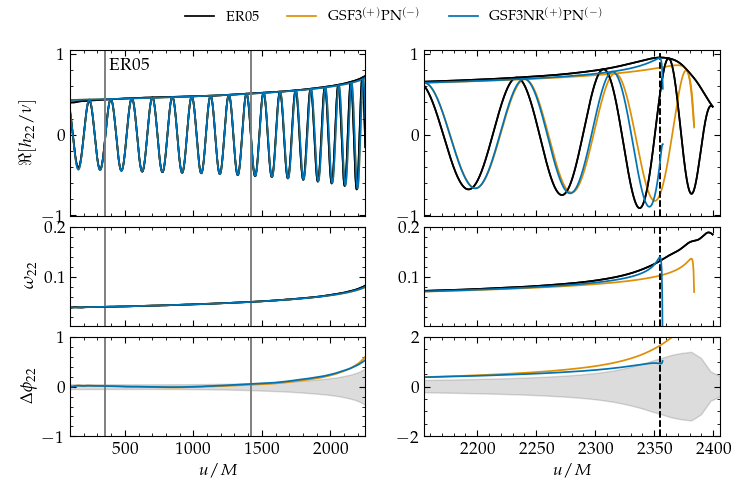}
  	\includegraphics[width=0.4\textwidth]{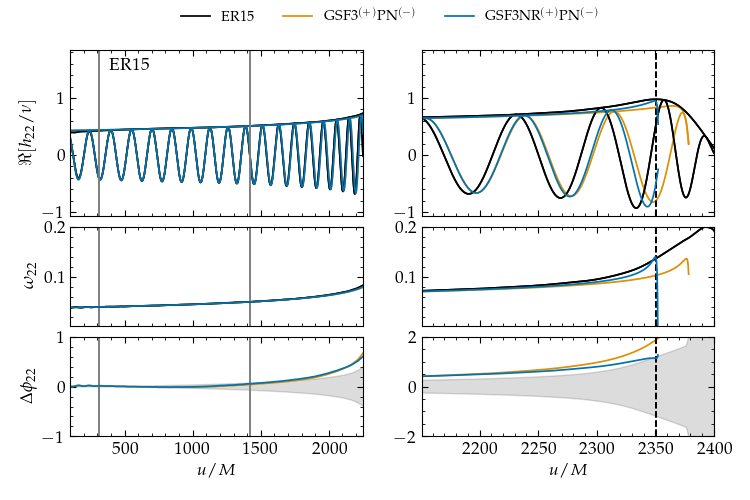}
  	\includegraphics[width=0.4\textwidth]{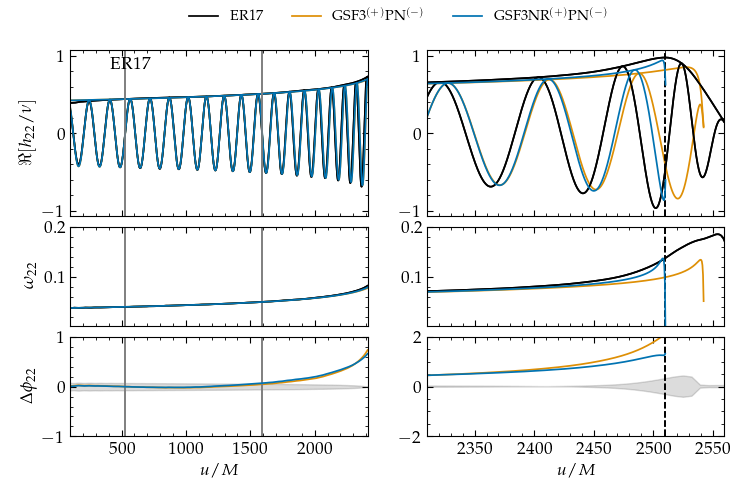}
  	\includegraphics[width=0.4\textwidth]{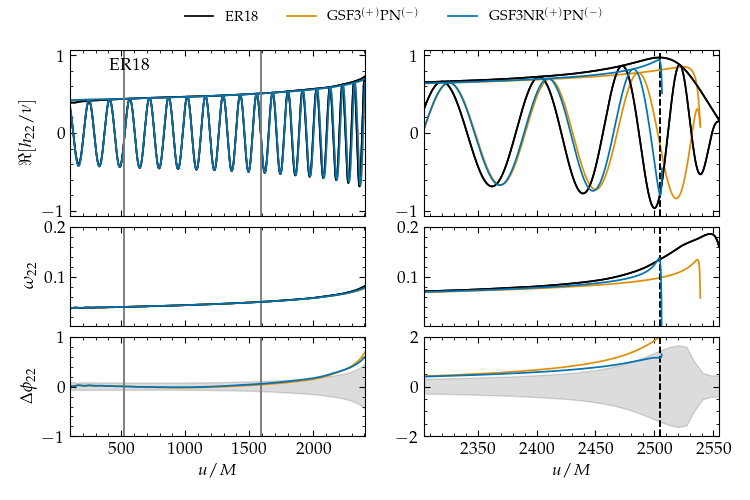}
	  \caption{Comparison between the two tidal flavors of the \TEOB~model, \gsftides{3} (orange lines) and \gsftides{3NR} (blue lines), 
	  and the eight $\tt H4$ unequal mass, 
	  spinning NR simulations summarized in Tab.~\ref{tab:nr_sims} (black lines). 
	  In each subfigure, the upper-left panel shows the \TEOB{} waveforms overlaid on top of the NR data during the early inspiral of the system. 
	  The upper-right panel shows instead the merger and few cycles preceding it. 
	  The middle and bottom panels, instead, show the frequency evolution and EOB/NR phase difference $\Delta\phi^{\rm EOBNR}$. 
	  The alignment region is delimited by two gray vertical lines, while merger is indicated with a dashed black line. 
	  In the bottom panels, the gray area denotes the NR error, estimated as described in App.~\ref{app:NR}.}
	\label{fig:bam_H4q1}
\end{figure*}

\begin{figure*}[t]
    \includegraphics[width=0.4\textwidth]{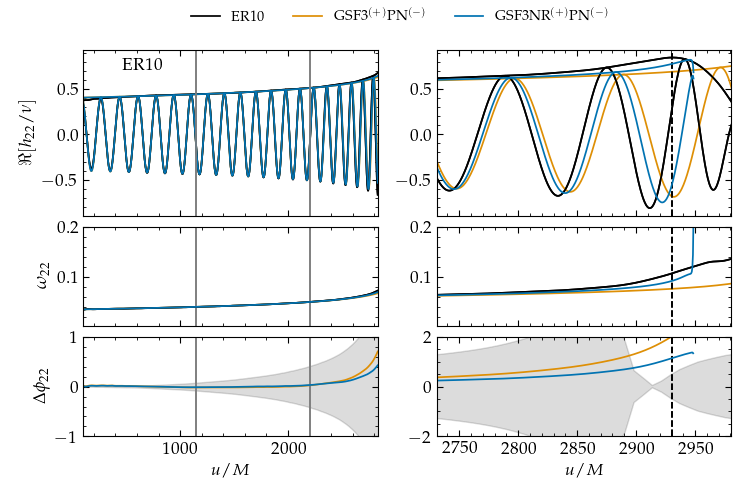}
    \includegraphics[width=0.4\textwidth]{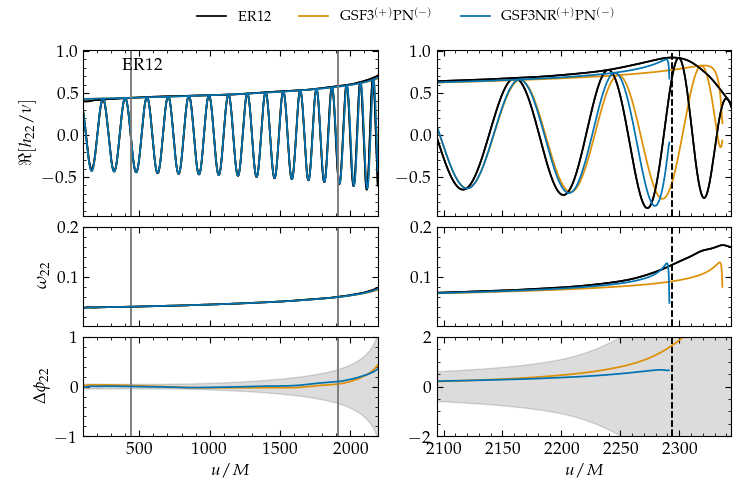}
    \includegraphics[width=0.4\textwidth]{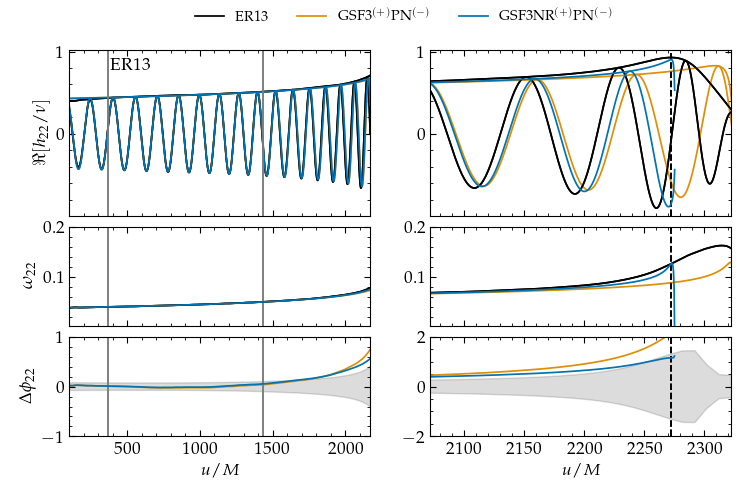}
    \includegraphics[width=0.4\textwidth]{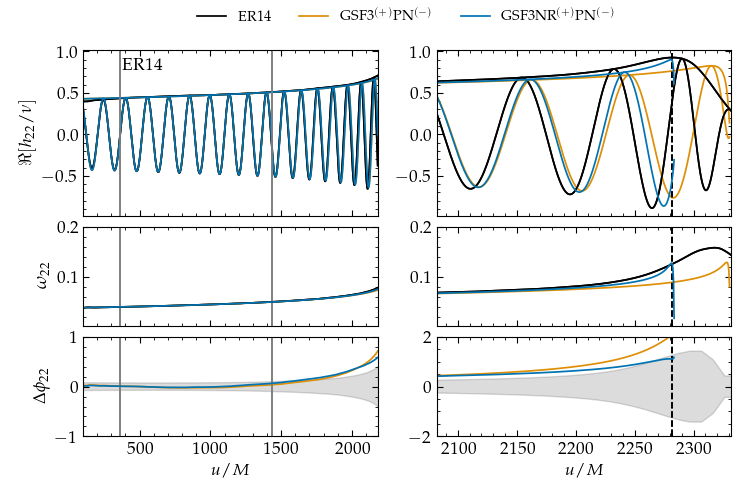}
	\caption{Same as Fig.~\ref{fig:bam_H4q1}, for the $\tt MS1b$ simulations of Tab.~\ref{tab:nr_sims}.}
	\label{fig:bam_MS1bq1}
\end{figure*}

To validate our new model, hereafter dubbed \gsftides{3NR}, in a regime far from the one considered in our calibration set, 
we align in the time domain the \gsftides{3} and the \gsftides{3NR} models 
with the NR simulations described in the previous sub-section over a window spanning the frequency range $\omega_{22}\in [0.4, 0.5]$.
When using the \gsftides{3NR} model, we iterate on the NQCs 5 times, employing them also in the flux, before performing the alignment.
The results of our alignment procedure can be inspected from Fig.~\ref{fig:bam_H4q1} and Fig.~\ref{fig:bam_MS1bq1} for the H4 and MS1b simulations, respectively.
Notably, we do not present results for the ER06 and ER11 simulations, as we were not able to correctly align them to the EOB model. 
Overall, we observe that although no unequal mass or spin-dependent corrections have been introduced to the \gsftides{3NR} model, it improves over
the previous one in all the comparisons presented.
The EOB/NR phase difference at merger is smaller than the estimated NR error in eight out of twelve cases displayed, with an average $\Delta\phi^{\rm EOBNR} \sim 0.5$ rad 
at merger. 
Unsurprisingly, the two cases for which significant disagreement -- larger than the NR error -- is found are both high $q$, large spins ones with the H4 EOS.
Notably, as was also the case for the simulations of Fig.~\ref{fig:bam_q1}, we observe again a faster increase of the NR frequency over the last two cycles with respect to the 
one predicted by the EOB model. This is especially evident in the $\tt ER05$, $\tt ER13$, $\tt ER14$, $\tt ER15$ and $\tt ER18$ simulations, where 
$\Delta\phi^{\rm EOB/NR}$ is smaller than the NR error in the inspiral and at merger, but larger than this quantity approaching the end of the coalescence.
This indicates that our model is not fully capturing the physics of the system beyond the contact of the two bodies.

\subsection{Unfaithfulness}
To further quantify the agreement of our model with NR,
we compute the EOB/NR unfaithfulness (or mismatch) $\bar{\mathcal{F}}$, which describes the global agreement
between our model and the NR data.
The EOB/NR unfaithfulness is defined as
\begin{equation}
\label{eq:fbar}
\bar{\mathcal{F}} = 1 - \max_{\phi_0, t_0} \frac{(h_{\rm EOB}, h_{\rm NR})}{\sqrt{(h_{\rm NR}, h_{\rm NR}) (h_{\rm EOB}, h_{\rm EOB})}}  \, ,
\end{equation}
where $(\cdot, \cdot)$ is the inner product in waveform space and $\phi_0, t_0$ are a reference time and phase. 
The action of the inner product on two generic waveforms $h$, $k$ is given by
\begin{equation}
(h, k) = 4\Re \int_{f_{\rm min}}^{f_{\rm max}} \frac{\tilde{h}(f) \tilde{k}^*(f)}{S_n(f)} df \, ,
\end{equation}
where $S_n(f)$ is the noise curve of the detector. For our comparisons, we employ the Einstein Telescope (ET) noise curve \cite{Hild:2010id},
and choose $f_{\rm min}$, $f_{\rm max}$ respectively as the initial frequency of the NR simulation (typically $> 300$ Hz) and the merger frequency. 

Applying Eq.~\eqref{eq:fbar}, we find that our model is either comparable with or improves the previous one in all 
but two cases, with typical mismatches below $1\%$.
Note that this number provides a conservative limit on the NR-faithfulness of our model, as little to no early inspiral
is included in our simulations. If hybrid waveforms were to be considered, the mismatch of \TEOB{} would decrease
as well.
This result confirms and complements the time domain phasing analysis discussed in the previous paragraph (see Tab.~\ref{tab:nr_sims}).

\section{Closed form representation of tidal sector}
\label{sec:tidal}

\begin{figure}[t]
	\includegraphics[width=0.5\textwidth]{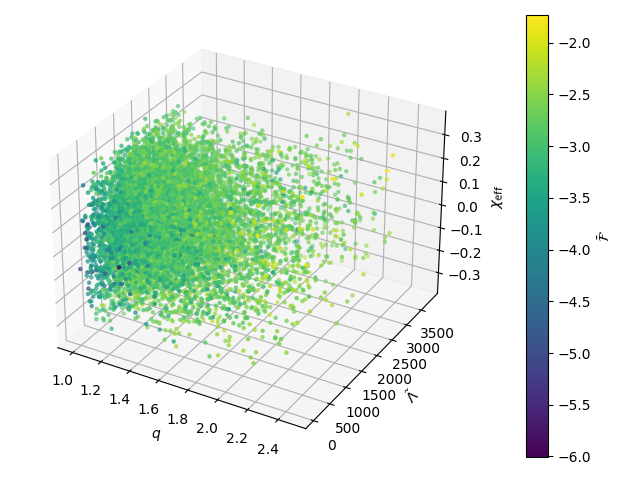}
	\caption{Mismatches between {\tt TEOBResumS} BBH baseline augmented by the phenomenological representation of phase and amplitude discussed 
	in Sec.~\ref{sec:tidal} and the full BNS model presented in this chapter. More that $99.9\%$ of the mismatches lie below $1\%$, 
	and $\sim 30\%$ below $1\text{\textperthousand}$}
	\label{fig:bam_q1}
\end{figure}

\begin{table*}
	\caption{
	The coefficients of $\Psi_{\Lambda}$ and $\Psi_{MQ}$ as defined in equations \ref{eq:phase:phenomodel}.
	}
	\label{tab:fits}
	% \resizebox{\textwidth}{!}{
		\begin{tabular}{c|cccc||cccc}
		\hline\hline
		  & \multicolumn{4}{c||}{$\Psi^{\Lambda}$} & \multicolumn{4}{|c}{$\Psi^{\rm MQ}$} \\
		\hline 
		& $d_1$ & $d_2$ &$n_{5/2}$& $n_3$ & $d_1$ & $d_2$ &$d_3$& $n_4$ \\
		\hline 
		$\nu=1/4$ & $99.80$ & $1560.60$ & $-2191.56$ & $5307.07$ & $-9.61$ & $23.12$ & $3.90$ & $-27.50$ \\
		$\nu\neq 1/4$ & $4.08$&$39.67$&$-86.62$& $158.92$ & $-2.84$&$28.93$& $63.34$&$-58.83$ \\
		\hline \hline
		\end{tabular}
	% }
\end{table*}

Following the ideas proposed in Ref.~\cite{Dietrich:2017aum} (see also
\cite{Kawaguchi:2018gvj,Dietrich:2018uni, Dietrich:2019kaq}), we develop 
a FD {\it closed-form} representation of the tidal sector of \TEOB{}
that can be employed  to augment any point-particle model of choice to
include the effects of tides. This representation is faithful ($\bar{\mathcal{F}} \lesssim \mathcal{O}(10^{-3})$) for $\Lambda\in[10,1000]$, mass
ratios $q\in[1,2.5]$ and spins $|\chi_i|\lesssim 0.05$. When widening the parameter range considered to $\Lambda\in[10,1000]$ and $|\chi_i|\lesssim 0.35$,
we find more that 99.9\% of the mismatches lie below 1\%, and $\sim30\%$ below $1\text{\textperthousand}$.

Phenomenological representations of tidal approximants are usually built from hybrid PN-EOB-NR
waveforms by (i) subtracing the point-mass (binary black hole) and
(ii) fitting the differences in phase and amplitude. The use of
numerical relativity data at high-frequencies potentially improves the
accuracy with respect to the ``exact'' (unknown) waveform, but also implies a 
limitation in the parameter space coverage.
For example, neither the {\tt NRTidal} model nor its improved version {\tt NRTidalv2} \cite{Dietrich:2019kaq} 
incorporate mass-ratio induced corrections, and take into account spin-quadrupole effects
only in the phase difference through the PN expression of~\cite{Nagar:2018plt}.
Moreover, the use of different approximants in the hybrid construction and in the
subtraction step can result in inconsistencies and systematic effects. 
The phenomenological representation of \TEOB{} does not have these
drawbacks, although it retains the uncertainties of \TEOB{} in the
merger description when compared to NR waveforms~\cite{Bernuzzi:2012ci,Bernuzzi:2015rla,Akcay:2018yyh}.

The phase of the $(2,2)$ FD waveform is modeled as the
sum of the contributions due to pure orbital interactions (O), 
pure tidal effects  ($\Lambda$), spin orbit and spin-spin effects (S),
and self-spin couplings, also known as monopole-quadrupole 
(MQ) terms. It formally reads
\be
\label{eq:phasetot}
\Psi(f) = \Psi_{\rm O} + \Psi_{\Lambda} + \Psi_{S} + \Psi_{\rm MQ}\,.
\ee
Under the simplifying assumption that these contributions can, indeed,
be clearly separated, the contribution to the phasing due to tidal
effects can be expressed as: 
\be
\label{eq:deltaphase}
\Delta\Psi(f) = \Psi^{\rm BNS}(f) - \Psi^{\rm BBH}(f) \approx \Psi_{\Lambda} + \Psi_{\rm MQ}\,.
\ee
The PN expression valid in the low-frequency, weak-field  regime of
$\Psi^{\Lambda}$ at incomplete 7.5PN accuracy was originally obtained 
in~\cite{Damour:2012yf}, and is of the form
\be
\label{eq:PsiFD}
\Psi_{\Lambda} = c_{\rm LO}^\Lambda x^{5/2}(1 + c_1^\Lambda x + c_{3/2}^\Lambda x^{3/2} + c_2^\Lambda x^2 + c_{5/2}^\Lambda x^{5/2})\, ,
\ee
while the self-spin contributions $\Psi_{\rm MQ}$ at 3.5PN accuracy 
is given by~\cite{Bohe:2015ana, Mishra:2016whh, Nagar:2018plt}
\begin{align}
\label{eq:PsiFDss}
\Psi_{\rm MQ} = \frac{3}{128 \nu} c_{\rm LO}^{\rm MQ} x^{-1/2}(1 + c_{1}^{\rm MQ, NLO} x + c_{3/2}^{\rm MQ, tail} x^{3/2})\,,
\end{align}
with the coefficients listed in Appendix~\ref{app:PNcoefs}. 
The functional form of our representations is obtained by 
the Pad\'e resummation of the PN expression, 
\begin{subequations}\label{eq:phase:phenomodel}
\begin{align}
\Psi_{\Lambda} =&  c_{\rm LO}^\Lambda x^{5/2}\frac{ 1 + \sum_{i=2}^{6}{n_{i/2}^\Lambda x^{i/2}}}{1 + \sum_{i=2}^{4}{d_{i/2}^\Lambda x^{i/2}}} \, ,\\
\Psi_{\rm MQ} =&  \frac{3}{128 \nu} c_{\rm LO}^{\rm MQ} x^{-1/2} \frac{ 1 + \sum_{i=2}^{3}{n_{i/2}^{\rm MQ} x^{i/2}}}{1 + \sum_{i=2}^{4}{d_{i/2}^{\rm MQ} x^{i/2}}}\,.
\end{align} 
\end{subequations}
The PN limit requires the following constraints on the pure tidal and spin-quadrupole coefficients:
\begin{subequations}
\label{eq:PadPhase}
\begin{align}
n_{1}^{\Lambda} =& c_1^{\Lambda} + d_1^{\Lambda} \, , \\
n_{3/2}^{\Lambda} =& (c_1^{\Lambda} c_{3/2}^{\Lambda} -  c_{5/2}^{\Lambda} - c_{3/2}^{\Lambda} d_1^{\Lambda} + n_{5/2}^{\Lambda})/c_1^{\Lambda} \, ,\\
n_{2}^{\Lambda} =& c_2^{\Lambda} + c_1^{\Lambda} d_1^{\Lambda} + d_2^{\Lambda} \, ,\\
d_{3/2}^{\Lambda} =& -(c_{5/2}^{\Lambda} + c_{3/2}^{\Lambda} d_1^{\Lambda} - n_{5/2}^{\Lambda})/c_1^{\Lambda} \, ,\\ 
n_1^{\rm MQ} =& c_1^{\rm MQ} + d_1^{\rm MQ}  \, ,\\
n_{3/2}^{\rm MQ} =& c_{3/2}^{\rm MQ} + d_2^{\rm MQ} \,.
\end{align} 
\end{subequations}
The remaining coefficients are fitted to \TEOB{PA}.
To incorporate corrections due to unequal-mass effects we parameterize
the free coefficients as a $\nu=1/4$ contribution plus a factor proportional to $\sqrt{1-4\nu}$. 
In particular, denoting a generic coefficient $n_i^{\Lambda}, d_i^{\Lambda}$ 
or $n_i^{\rm MQ}, d_i^{\rm MQ}$ as $p_i^{\Lambda}$ 
and $p_i^{\rm MQ}$, we have:
\begin{subequations}
\begin{align}
p_i^{\Lambda} =& p_i^{(\nu=1/4)} + \frac{\kappa_1 - \kappa_2}{\kappa_1 + \kappa_2}p_i^{(\nu \neq 1/4)}\sqrt{1-4\nu} \, , \\
p_i^{\rm MQ}  =& p_i^{(\nu=1/4)} + \frac{(C_{\rm Q1} \tilde{a}_1^2 - C_{\rm Q2}\tilde{a}_2^2)}{(C_{\rm Q1} \tilde{a}_1^2 + C_{\rm Q2}\tilde{a}_2^2)}p_i^{(\nu \neq 1/4)}\sqrt{1-4\nu} \, .
\end{align} 
\end{subequations}
These functional forms are inspired by the Taylor expansions of the
coefficients known from PN theory about $\nu=1/4$. 

We then proceed as follows: (i) we compute the phase difference between a
set of $\nu=1/4$, nonspinning waveforms, and obtain the values of $(d_1^{\Lambda}, d_2^{\Lambda}, n_{5/2}^{\Lambda}, n_3^{\Lambda})^{(\nu=1/4)}$; (ii)
we consider a set of unequal mass, nonspinning waveforms and fit $(d_1^{\Lambda}, d_2^{\Lambda}, n_{5/2}^{\Lambda}, n_3^{\Lambda})^{(\nu \neq 1/4)}$,
setting the $\nu=1/4$ coefficients to the values found in the previous point; (iii) we fit
equal mass, spinning waveforms, and find the values of $(d_1^{\rm MQ}, d_2^{\rm MQ}, n_4^{\rm MQ}, d_3^{\rm MQ})^{(\nu=1/4)}$ using
again the equal-mass coeffcients of (i); (iv) we use all information
found up to now, and fit unequal mass, spinning waveforms to find
$(d_1^{\rm MQ}, d_2^{\rm MQ}, n_4^{\rm MQ}, d_3^{\rm MQ})^{(\nu \neq 1/4)}$.  
The values of all fitted coefficients, obtained from a dataset of $\sim 1000$ waveforms, are summarized in
Table~\ref{tab:fits}. 

We proceed analogously for the amplitude, whose
tidal and self-spin terms are modeled as
\begin{align}
\label{eq:AmpTidal}
\tilde{A}^\Lambda =& \sqrt{\frac{2 \nu}{3}} \pi x^{13/4} c_{5}^\Lambda \frac{1 + \frac{c_{6}^\Lambda}{c_{5}^\Lambda} x + \frac{ 22672}{9} x^{2.89}}{1 + d_4^\Lambda x^4} \, ,\\
\label{eq:AmpSpin}
\tilde{A}_{\rm MQ} =&-\sqrt{\frac{3 \nu}{2}} \pi x^{1/4}\frac{(C_{Q1} \tilde{a}_1^2 + C_{Q2} \tilde{a}_2^2)}{1 + e_4 x^4} \,.
\end{align}
In Appendix~\ref{app:PNcoefs} we compute the monupole-quadrupole interactions to the
FD amplitude, that are used to constrain some of the fitting coefficients.
In Eq.~\eqref{eq:AmpSpin} we incorporate only LO PN information 
for the spin sector, as we find that, while the addition of the NLO term slightly improves
the low frequency behavior of our fits, it also negatively impacts
the overall agreement in the high-frequency regime.
The unknown coefficients are fit by following the same procedure of the previous paragraph. We find:
\begin{subequations}
\begin{align}
d_4^{(\nu=1/4)} =& 5009.8736694 \, ,\\
d_4^{(\nu\neq 1/4)}=& -4017.88863642 \, ,\\
e_4^{(\nu=1/4)} =&  5.98351934 \, ,\\
e_4^{(\nu \neq 1/4)} =& 20.04283392 \,.
\end{align}
\end{subequations}

We compute the unfaithfulness between BNS {\tt TEOBResumSPA} waveform and the BBH
{\tt TEOBResumSPA} model 
augmented by the phenomenological description of tides.
We consider $10^4$ systems with masses uniformly distributed between $[1, 2.5]$ $M_{\odot}$, dimensionless spins uniformly distributed in $[-0.35, 0.35]$ and tidal 
parameters in $[10, 3000]$.
When restricting the calculation to non-spinning binaries with $\tilde{\Lambda} < 1000$ and $m_{1,2} \in [1, 2.5] \Msun$ 
we find a maximum unfaithfulness of $2 \times 10^{-3}$. Widening the range of tidal parameters, we find that the 
faithfulness degrades for tidal deformabilities larger than roughly $\sim 2000$.  
The worst matches ($\bar{\mathcal{F}} \sim 2\%$)
are obtained, as expected, for unequal mass configurations with large $\tilde\Lambda$.
When also considering spins, we find that the largest differences are obtained with configurations having at least one rapidly spinning NS.
In this scenario, unfaithfulness values can increase above the nominal $1\%$ threshold.
When the spins of the NS are moderate ($|\chi_{A,B}|< 0.05$), instead, we find mismatch values around $\mathcal{O}(10^{-3})$ or lower.

\section{Conclusions}
\label{sec:conc}

In this paper we presented a new, improved gravitational wave model for coalescing binary neutron stars.
Building on a previous version of \TEOB{}, we included: (i) leading order magnetic tidal corrections to the 
$B$ EOB potential; (ii) 2PN electric tidal terms in the $(\ell,m) = \{(2,2), (2,1), (4,2), (4,4) \}$ waveform multipoles;
(iii) 5.5PN spin-orbit terms in the gyro-gravitomagnetic coefficients entering the EOB Hamiltonian; (iv) NQC corrections to the quadrupolar $(2,2)$ mode; (v) a new NR-informed parameter, $\alpha$, which enters the GSF-resummed $A$ tidal potential and is fit against four high resolution NR simulations from the CoRe database.
The model was then benchmarked against 14 new eccentricity reduced BNS simulations for spinning binaries, here presented for the first time. 
The new tidal \TEOB{} improves over the previous model in terms of both mismatches and time-domain phasing, 
representing a robust and efficient alternative to current state of the art tidal models.
Finally, we presented a frequency domain closed form representation of the matter contributions to the phase and amplitude of the quadrupolar (2,2) 
mode implied by our full EOB model. This phenomenological representation faithfully approximates \TEOB{} ($\bar{\mathcal{F}} \lesssim 0.001$) 
over a considerable portion of the parameter space, with the worst mismatches obtained for strongly asymmetric systems with large spins.

In spite of the model's performance being overall satisfactory for many of the systems we inspected, 
our comparisons highlighted how improvements are necessary in the last few cycles before merger. 
Comparisons between our model and NR waveforms highlight
that the frequency evolution of the NR waveforms close to merger is steeper with respect to the one provided by 
the models. This indicates the need of further improving the treatment of matter effects beyond contact, which will
require large amounts of sufficiently accurate NR data, spanning large portions of the BNS parameter space. 
This ambitious goal will enable precise inference of the equation of state of cold, dense matter with XG
detectors such as ET \cite{Maggiore:2019uih} and Cosmic Explorer (CE) \cite{Reitze:2019iox}.

\begin{acknowledgments}
  We thank Thibault Damour for fruitful discussions during the preparation of this manuscript
  and for	partly supporting this work via the ``2021 Balzan Prize for Gravitation: 
  Physical and Astrophysical Aspects''.
  RG is supported by the Deutsche Forschungsgemeinschaft (DFG) under Grant No.
  406116891 within the Research Training Group RTG 2522/1. 
  SB acknowledges support by the EU Horizon under ERC Consolidator Grant, no. InspiReM-101043372.
  NO acknowledges support by the UNAM-PAPIIT Grant No. IA101123.
  WT acknowledges support by the National Science Foundation under grant PHY-2136036.
  The authors acknowledge the Gauss Centre for Supercomputing e.V. for
  funding this project by providing computing time on the
  GCS Supercomputer SuperMUC-NG at LRZ (allocations
  $\tt pn36ge$ and $\tt pn36jo$).
  \TEOB{} is publicly available at
 
  {\footnotesize \url{https://bitbucket.org/eob_ihes/teobresums/src/master/}}

\end{acknowledgments}

\clearpage
\widetext
\appendix

\section{GSF coefficients of the $A$ EOB potential}
\label{app:APN}

For completeness, we collect here the %PN (up to NNLO) and 
GSF coefficients entering the radial
potential $A(r)$, see Sec.~\ref{sec:eobmodel}.
%\subsection{PN coefficients}

For $\ell=2$, the 0GSF, 1GSF, 2GSF terms explicitly read:
\begin{eqnarray}
\label{eq:A2p_gsf}
\hat{A}_A^{(2+)\rm 0GSF} =& 1 + \frac{3 u^2}{1-\rLR u} \, , \\
\hat{A}_A^{(2+)\rm 1GSF} =& \frac{1}{(1-\rLR u)}\frac{5}{2} u (1-a_1 u)(1-a_2 u) \frac{1+n_1 u}{1 + d_2 u^2} \, , \\
\hat{A}_A^{(2+)\rm 2GSF} =& \frac{337}{28} \frac{u^2}{(1-\rLR u)^p} \, ,
\end{eqnarray}
while for $\ell=3$ we have:
\begin{eqnarray}
\label{eq:A3p_gsf}
\hat{A}_A^{(3+)\rm 0GSF} =& (1-2u)(1+\frac{8}{3}\frac{u^2}{1-\rLR u}) \, , \\
\hat{A}_A^{(3+)\rm 1GSF} =& \frac{1}{(1-\rLR u)^{7/2}}\frac{15}{2} u (1+c_1 u +c_2u^2+c_3 u^3) \\
& \times \frac{1+c_4 u + c_5 u^2}{1 + c_6 u^2} \, , \\
\hat{A}_A^{(3+)\rm 2GSF} =& \frac{110}{3} \frac{u^2}{(1-\rLR u)^p} \, .
\end{eqnarray}
The parameters ($a_1, \dots,a_4$) and ($c_1, \dots,c_6$) can be read from Eq. (17) and Eq. (29) of \cite{Akcay:2018yyh}.

\section{Simulations}
\label{app:NR}

\begin{table}
\begin{tabular}{ccccccccc}
\hline\hline
Sim. & $L$ & $l^{\rm mv}$ & $n$ & $n^{\rm mv}$ & $h_{L-1}$ & $h_0$ \\
\hline 
ER01 & 8 & 2 & [144,216,288] & [72,108,144] & [0.25,0.16667,0.125] & [32.0,21.33376,16.0]\\
ER02 & 7 & 2 & [128,192,256] & [64,96,128] & [0.28125,0.1875,0.140625] & [18.0,12.0,9.0]\\
ER03 & 7 & 2 & [128,192,256] & [64,96,128] & [0.28125,0.1875,0.140625] & [18.0,12.0,9.0]\\
ER04 & 8 & 2 & [144,216,288] & [72,108,144] & [0.25,0.16667,0.125] & [32.0,21.33376,16.0]\\
ER05 & 8 & 2 & [144,216,288] & [72,108,144] & [0.25,0.16667,0.125] & [32.0,21.33376,16.0]\\
ER06 & 7 & 2 & [128,192,256] & [64,96,128] & [0.249,0.166,0.1245] & [15.936,10.624,7.968]\\
ER10 & 7 & 2 & [128,192,256] & [64,96,128] & [0.28125,0.1875,0.140625] & [18.0,12.0,9.0]\\
ER11 & 7 & 2 & [128,192,256] & [64,96,128] & [0.249,0.166,0.1245] & [15.936,10.624,7.968]\\
ER12 & 7 & 2 & [128,192,256] & [64,96,128] & [0.28125,0.1875,0.140625] & [18.0,12.0,9.0]\\
ER13 & 7 & 2 & [128,192,256] & [64,96,128] & [0.28125,0.1875,0.140625] & [18.0,12.0,9.0]\\
ER14 & 7 & 2 & [128,192,256] & [64,96,128] & [0.28125,0.1875,0.140625] & [18.0,12.0,9.0]\\
ER15 & 7 & 2 & [128,192,256] & [64,96,128] & [0.28125,0.1875,0.140625] & [18.0,12.0,9.0]\\
ER17 & 7 & 2 & [128,192,256] & [64,96,128] & [0.28125,0.1875,0.140625] & [18.0,12.0,9.0]\\
ER18 & 7 & 2 & [128,192,256] & [64,96,128] & [0.28125,0.1875,0.140625] & [18.0,12.0,9.0]\\
\hline\hline
\end{tabular}
\caption{Grid parameters for the NR simulations. From the second to the last column: Refinement levels, number of moving levels (finest), number of points per direction in non-moving refinement levels, number of points per directions in moving refinement levels, grid spacing per direction in the finest level, grid spacing in the coarser level.}
\label{tab:nr_grids}
\end{table}

\begin{figure}[t]
    \includegraphics[width=0.45\textwidth]{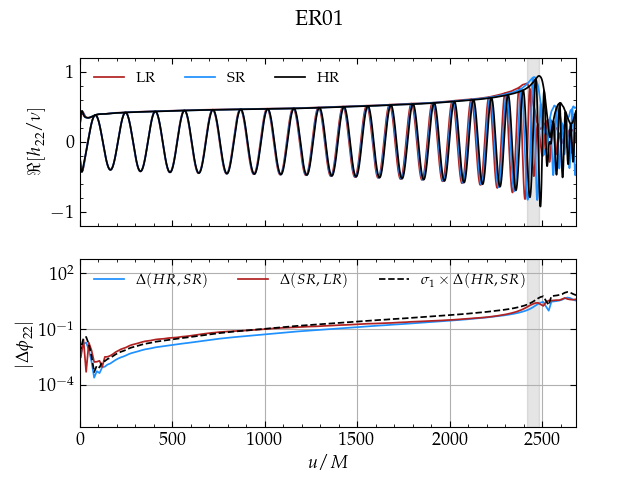}
    \includegraphics[width=0.45\textwidth]{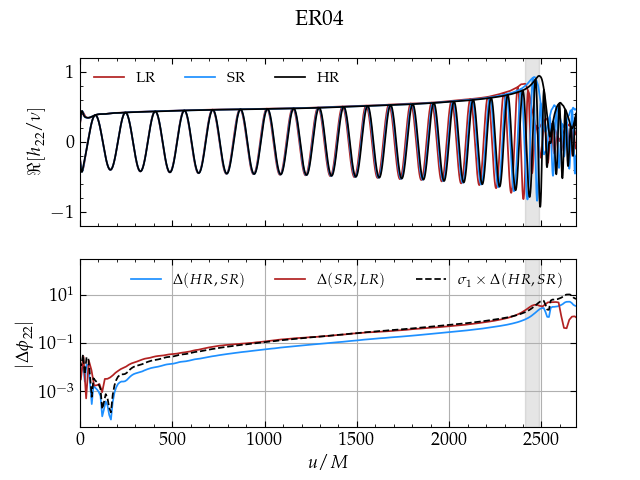}
    \includegraphics[width=0.45\textwidth]{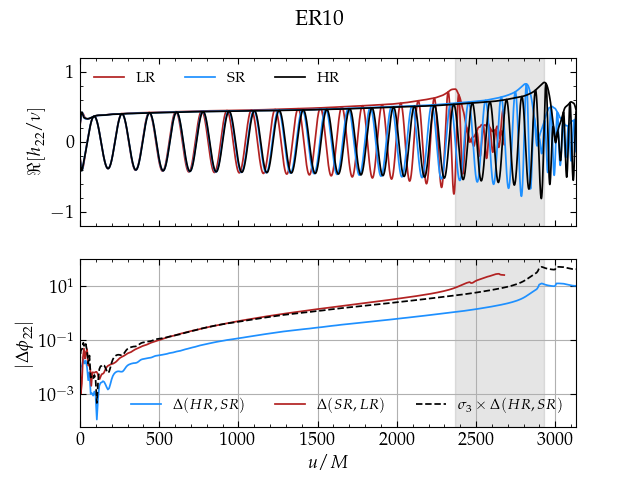}
    \includegraphics[width=0.45\textwidth]{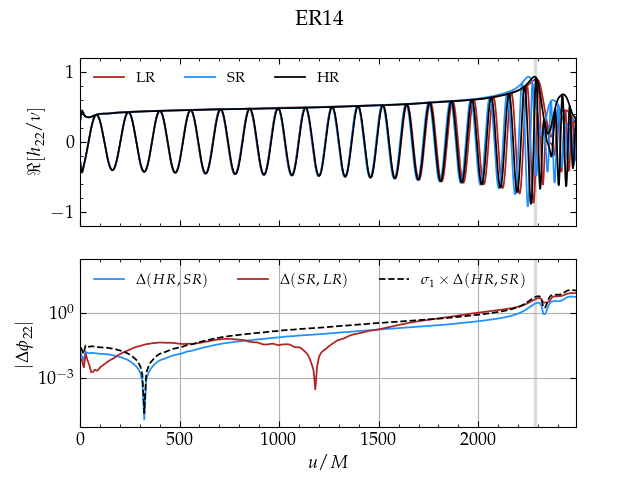}
	\caption{Convergence plots for four sample simulations presented in this work, two for each of the EOS considered. 
			We display the real part of the $h_{22}$ mode (top panels) 
			and the phase differences between resolutions (bottom panels), as well as the 
			rescaled phase difference assuming first ($\tt ER01$, $\tt ER04$, $\tt ER14$) or third ($\tt ER10$) 
			order convergence.
			}
	\label{fig:convplots}
\end{figure}

Initial data for the simulations are prepared using the formalism and iterative procedure described in Refs.~\cite{Moldenhauer:2014yaa,Dietrich:2015pxa}. 
Typical initial eccentricities estimates from the coordinate distance are of the order of $10^{-4}$ after four iterations.
Simulations are performed with the methods described in Ref.~\cite{Bernuzzi:2016pie}, in particular a high-order scheme for 
hydrodynamics based on characteristic reconstruction and the WENOZ reconstruction. 
The grid setups employed in the simulations are detailed in Tab.~\ref{tab:nr_grids}.
Their self-convergence is assessed by computing the phase difference between $h_{22}$ considered at different resolutions.
The convergence rate $p$ is found by rescaling such differences by the scaling factor $\sigma_p$, defined as~\cite{Gonzalez:2022mgo}
\be
\sigma_p = \frac{\Delta^p_{\rm LR} - \Delta^p_{\rm SR}}{\Delta^p_{\rm SR} - \Delta^p_{\rm HR}} \, ,
\ee
where (LR, SR and HR) denote respectively the low, standard and high resolution simulations and $\Delta_x$ is the 
grid spacing at the resolution $x$.
Figure~\ref{fig:convplots} shows the self convergence test for four representative simulations.
Overall, we find that ten of the new simulations behave like $\tt ER01$ or $\tt ER14$ shown, 
which fail to show clear convergence over the full waveform. 
The remaining four simulations, instead, converge either at first or third order, similar to $\tt ER04$ or $\tt ER10$ shown.

Given the lack of clear convergence for most of the simulations considered,
we do not Richardson extrapolate the waveforms. We do, however, extrapolate 
them to null infinity using a polynomial in $1/R$ of order $K=2$.
The error budget of the simulation is then estimated as the sum in quadrature of the resolution error,
computed considering the phase difference between the highest and second highest simulations, and the extraction error,
computed as the phase difference between the waveform extrapolated to null infinity and the waveform extracted at the largest $R$
available (typically either 1000M or 1600M).

\section{Fit for $\dot{\omega}_{\rm mrg}^{\rm NR}$}
\label{app:domg_fit}
We fit the time derivative of the NR frequency at merger employing all simulations associated to the 
first release of the {\tt CoRe} database \cite{Dietrich:2018phi} and a handful of simulations from the {\tt SACRA} database \cite{Kawaguchi:2018gvj,Kiuchi:2017pte,Kiuchi:2019kzt}.
Following Ref.~\cite{Breschi:2022xnc}, we assume a functional form of the type:
\begin{equation}
\dot{\omega}_{\rm mrg}^{\rm NR} = a_0 Q^M(X) Q^S(\hat{S}, X) Q^T(\kappa^T_2, X) \, ,
\label{eq:domg_mrg_fit}
\end{equation}
where $X=X_A - X_B$, $\hat{S} = S_A + S_B$, $\kappa_2^T = \kappa_A^{(2+)} + \kappa_B^{(2+)}$, and:
\begin{align*}
Q^M &= 1+ a_1^M X \, ,\\
Q^S &= 1+ a_1^S(1+b_1^S X) \hat{S} \,, \\
Q^T &= \frac{1 + a_1^T(1+b_1^T X)^T \kappa_2 ^T + a_2 ^T(1+b_2 ^T X){\kappa_2^T}^2}{1 + a_3^T(1+b_3^T X)\kappa_2^T + a_4^T(1+b_4^T X) {\kappa_2^T}^2} \, .
\end{align*}
The coefficients of the fit can be read from Tab.~\ref{tab:domg_fits}.

\begin{table}
	\centering
	% \resizebox{\textwidth}{!}{
	  \begin{tabular}{c|cccc||cccc}
	  \hline\hline
	  X  &  $a_1^X$ & $a_2^X$ &$a_3^X$& $a_4^X$ & $b_1^X$ & $b_2^X$ &$b_3^X$& $b_4^X$ \\
	  \hline 
	  $M$ & -1.7988 & -- &--      & --     & --      & --  & --     & -- \\
	  $S$ & 0.3555  & -- &--      & --     & -7.1674 & --  & --     & -- \\
	  $T$ & 0.0314  & 0  & 0.1714 & 0.0006 & -6.8144 & 1.0 & 5.1651 & -1.0333 \\
	  \hline \hline
	  \end{tabular}
	% }
	\caption{  
	\label{tab:domg_fits}
	The coefficients for $\dot{\hat{\omega}}_{\rm mrg}^{\rm NR}$ from Eq.~\ref{eq:domg_mrg_fit}. The fitted value of $a_0$ is $a_0=0.0074$.
	}
\end{table}

\section{Tidal PN coefficients for phase and amplitude}
\label{app:PNcoefs}

We collect the PN coefficients entering the
phenomenological representation of \TEOB{}. We also compute for the
first time - to our knowledge - 
the contribution of spin-quadrupole interactions to the frequency
domain amplitude $\tilde{A}_{22}$. 

\subsection{Phase}
The pure tidal PN coefficients entering Eq. \eqref{eq:PsiFD} and
Eq. \eqref{eq:PadPhase} beyond leading order (LO) can be compactly 
expressed as 
\be 
c_i^{\Lambda} = \Bigl[-\kappa_A^{(2+)} \frac{3}{16 \nu} \frac{(12 - 11 X_A)}{(1 - X_A)} \bar{c}^\Lambda_i(X_A) + (A \leftrightarrow B)\Bigr]/c^{\Lambda}_{\rm LO} \,,
\ee
where $i$ denotes the relative PN order beyond LO, and the $c^{\Lambda}_{\rm LO}, \bar{c}^\Lambda$ coefficients
are given by:

%\begin{widetext}
\begin{subequations}
\begin{align}
c^\Lambda_{\rm LO}    =& -\kappa_A^{(2+)} \frac{3}{16 \nu} \frac{(12 - 11 X_A)}{(1 - X_A)} + (A \leftrightarrow B) \, ,\\
\bar{c}^\Lambda_1     =& -\frac{2}{4} \frac{5(260 X_A^3 - 2286 X_A^2 - 919 X_A + 3179) }{336 (11 X_A - 12)} \, ,\\
\bar{c}^\Lambda_{3/2} =& -\pi \, ,\\
\bar{c}^\Lambda_2     =& \frac{5 (67702048 X_A^5 - 223216640 X_A^4 + 337457524 X_A^3 - 141992280 X_A^2 + 96008668 X_A - 143740242)}{3 \times 3048192 (11 X_A - 12)} \, ,\\
\bar{c}_{5/2} =& -\frac{\pi (10232 X_A^3 - 7022 X_A^2 + 22127 X_A - 27719)}{192 (11 X_A - 12)} \,.
\end{align}
\end{subequations}
%\end{widetext}
The monupole-quadrupole interactions included in Eq.~\eqref{eq:PsiFDss} instead read:
%\begin{widetext}
\begin{subequations}
\begin{align}
c_{\rm LO}^{\rm MQ}  =& -50 (C_{QA} \tilde{a}_A^2 + C_{QB} \tilde{a}_B^2) \, ,\\
c_{1}^{\rm MQ}  =& \frac{5}{84}\Bigl[(9407 + 8218 X_A + 2016 X_A^2) C_{QA}  \tilde{a}_A^2 + (A  \leftrightarrow B)\Bigr]/c_{\rm LO}^{\rm MQ} \, ,\\
c_{3/2}^{\rm MQ} =& \Bigl\{ 10\Bigl[(X_A + 308/3)\tilde{a}_A - (X_A + 86/3)\tilde{a}_B - 40 \pi\Bigr] C_{QA} \tilde{a}_A^2 - 440 C_{\rm OcA} \tilde{a}_A^3 +  (A  \leftrightarrow B) \Bigr\} /c_{\rm LO}^{\rm MQ} \,,
\end{align} 
\end{subequations}
%\end{widetext}
where $\tilde{a}_i = X_i \chi_i$, $i=A,B$.

\subsection{Amplitude}
The SPA allows us to compute the corrections to the frequency domain
amplitude of the waveform beyond leading PN order. Assuming that the
only modes contributing to the waveform are those with $\ell=2, |m|=2$
and denoting the amplitude of the 22 mode as $A_{22}$, one
has that
\begin{subequations}
\begin{align}
\label{eq:af_def}
\tilde{A}_{22} = & A_{22}(t_f)\sqrt{\frac{\pi}{\ddot{\phi}(t_f)}} \, ,\\
\ddot{\phi}(x) = & -\frac{3}{2} x^{1/2} \frac{\mathcal{F}(x)}{E'(x)} \,.
\end{align}
\end{subequations}
Where $E(x)$ and $\mathcal{F}(x)$ denote, respectively, the energy and its flux.
We employ expressions for $A_{22}$, $E(x)$ and $\mathcal{F}(x)$ 
which contain point-particle and spin-orbit corrections known up to 3.5 PN
\cite{Mikoczi:2005dn, Arun:2008kb, Buonanno:2009zt}, spin-quadrupole corrections up to
relative 1PN \cite{Bohe:2015ana, Messina:2018ghh}, and pure tidal corrections up to relative 1PN \cite{Vines:2011ud, Damour:2012yf, Wade:2014vqa}.
Expanding Eq. \eqref{eq:af_def}, we find that the waveform amplitude
is given by 
\be
\label{af_gen}
\tilde{A}_{22}(x) = \sqrt{\frac{2 \nu}{3}}\pi x^{-7/4} \sum_{i=0}^{i=X}{c_{i/2}   x^{i/2}} \,.
\ee
The coefficients of the point mass and spinnning terms are given in
e.g \cite{Khan:2015jqa}. Spin-quadrupole and tidal effects enter the amplitude at
relative 2 and 5 PN orders. The LO and NLO coefficients of both are given by: 
\begin{subequations}
\begin{align}
\label{af_coeff}
c_2^{\rm MQ} =& -\frac{3}{2}(C_{QA} \tilde{a}_A^2 + C_{QB}\tilde{a}_B^2) \, ,\\
c_{5/2}^{\rm MQ} =& -C_{QA} \tilde{a}_A^2 (\frac{12247}{1344}-\frac{2221}{672} \sqrt{1-4\nu}+\frac{53}{336} \nu ) +
      -C_{QB} \tilde{a}_B^2 (\frac{12247}{1344}+\frac{2221}{672} \sqrt{1-4\nu}+\frac{53}{336} \nu ) \, ,\\
c_{5}^{\Lambda} =& \frac{27}{2} \nu \left[(\Lambda_A + \Lambda_B)(-1 + 3\nu) + (\Lambda_A - \Lambda_B)\sqrt{1 - 4 \nu} (\nu -1)\right] \, ,\\
c_{6}^{\Lambda} =& \frac{1}{448}\Bigl[(\Lambda_A+\Lambda_B)(930 -40455 \nu + 131233 \nu^2 - 37748 \nu^3) 
     + (\Lambda_A - \Lambda_B)\sqrt{1-4\nu}(930 - 38595 \nu + 55903 \nu^2 
		        + 588 \nu^3)\Bigr] \,.
\end{align}
\end{subequations}
\end{document}